\documentclass[a4paper,12pt]{article}

\usepackage{geometry}
\usepackage{xcolor}
\usepackage{hyperref}
\usepackage{amsmath,amssymb}
\usepackage[capitalise]{cleveref}
\usepackage{graphicx}

\usepackage{parskip}

\usepackage[T1]{fontenc}
\usepackage[utf8]{inputenc}
\usepackage[scaled=1.04]{biolinum}

\usepackage{fourier}
\usepackage[cal=stixplain]{mathalpha}
\usepackage[scaled=0.83]{beramono}
\usepackage{microtype}

\usepackage{titlesec}
\titleformat{\section}{\sffamily\Large\bfseries}{\sffamily\Large\bfseries\thesection}{0.5em}{\sffamily\Large\bfseries}
\titleformat{\subsection}{\sffamily\large\bfseries}{\sffamily\large\bfseries\thesubsection}{0.5em}{\sffamily\large\bfseries}

\newcommand{\linkemail}[1]{\href{#1}{\texttt{#1}}}

\usepackage{natbib}
\usepackage[english]{babel}

\definecolor{links}{rgb}{0.26,0.41,0.88}
\hypersetup{
    colorlinks=true,
    allcolors=links,
}

\usepackage[style=english]{csquotes}

\usepackage{etoolbox}

\makeatletter
\newcommand\bibstyle@and{\bibpunct(){ and }a,,}
\newcommand\bibstyle@semicolon{\bibpunct();a,,}
\newcommand\bibstyle@comma{\bibpunct(),a,,}
\makeatother

\newcommand{\citeand}[2]{\citestyle{comma}\cite{#1} and \cite{#2}\citestyle{semicolon}}

\newcommand{\possessivecite}[1]{\textcolor{links}{\citeauthor{#1}'s} \citeyearpar{#1}}
\newcommand{\possessivemulticite}[2]{\textcolor{links}{\citeauthor{#1}'s} \citeyearpar{#1,#2}}

\newcommand{\given}{\,|\,}

\begin{document}

\thispagestyle{empty}
\renewcommand{\thefootnote}{\fnsymbol{footnote}}

\begin{center} 
{\Large\bf\sffamily It's all in your head --- fine-tuning arguments do not require aleatoric uncertainty}
\end{center}
\vspace{3mm}
\begin{quote}
\begin{center}
{\begin{large}\textsf{\textbf{Andrew Fowlie}}\end{large}\thanks{\textsf{\linkemail{andrew.fowlie@xjtlu.edu.cn}}}}\\[6mm]
{\textit{X-HEP Laboratory, Department of Physics,\\ School of Mathematics and Physics, Xi'an Jiaotong-Liverpool University, Suzhou, 215123, China}\\}
\end{center}
\end{quote}
\begin{quote}
Prompted by misconceptions in the recent literature, we review the justifications for naturalness arguments and Occam's razor found in Bayesian statistics. We discuss the automatic Occam's razor that emerges in Bayesian formalism, bringing together points of view from diverse fields, including statistics, social sciences, physics and machine learning. In pedagogical calculations, we demonstrate that this automatic razor disfavors unnatural models in which predictions must be fine-tuned to agree with observation.
\end{quote}












\makeatletter
\@thanks
\makeatletter
\renewcommand{\thefootnote}{\arabic{footnote}}
\setcounter{footnote}{0}

\section{Introduction}

Naturalness is an important yet contentious idea in theoretical physics. The idea posits that we may judge theories by whether their predictions must be fine-tuned to agree with experimental observations. Theories that require fine-tuning are considered unnatural. This led to a search for natural theories that dominated model-building in particle physics since the 1980s~\citep[see e.g.,][]{giudice2008,craig2023,peskin2025}.

Theories motivated by naturalness, such as supersymmetry~\citep[see e.g.,][]{martin1998}, typically predicted new particles that could be within reach of the Large Hadron Collider~(LHC) or its predecessors. The failure to observe such particles led to criticism \citep{richter2006,hossenfelder2019} and re-evaluation of the idea of naturalness~\citep{feng2013,dine2015,wells2019a,wells2019b,wells2025}. At the same time,  connections were established between Bayesian inference and naturalness. This was based on observations that Bayesian inference contains an automatic Occam's razor that penalized unnatural theories~\citep{jaynes1979,jefferys1991,jefferys1992a,jefferys1992b,mackay1992a,mackay1992b}. 

We seek to clarify a misconception about the nature of probability in recent discussions on fine-tuning~\citep{hossenfelder2019,wells2025}. Specifically, we emphasize that the probabilities in fine-tuning arguments are \emph{epistemic}, that is, related to a rational degree of belief, and that fine-tuning arguments do not assume or require \emph{aleatoric} uncertainties or randomness. The clarification draws on textbook material on statistical inference that is not controversial in that setting, though it has been subject to historical controversy. To make this point, we begin by defining epistemic and aleatoric uncertainty in \cref{sec:uncertainty}; we continue in \cref{sec:occam} by discussing the automatic Occam's razor that penalizes unnatural theories and in \cref{sec:naturalness} connect it to naturalness in theoretical physics. We conclude in \cref{sec:discuss} with a critical discussion of \citeand{hossenfelder2019}{wells2025}.

\section{Uncertainty}\label{sec:uncertainty}

There are many things that we don't know; we are uncertain about those things. Of course, you may know more than I do, and so uncertainty must be personal and subjective. The things we are uncertain about might be in the future, such as the weather tomorrow, or in the past, such as the weather the day you were born, or they might be well-known by some people but unknown to us, such as the capital of Liberia~\citep{lindley2006}. We may describe our uncertainties using probability theory. In this setting, probabilities represent our rational degree of belief.

This led to \cite{definetti2017} to famously declare that probabilities do not exist in an objective sense:

\begin{displaycquote}[][capitals in original]{definetti2017}[]
My thesis, paradoxically, and a little provocatively, but nonetheless genuinely, is
simply this:
\begin{center}
PROBABILITY DOES NOT EXIST  
\end{center}
The abandonment of superstitious beliefs about the existence of Phlogiston, the Cosmic
Ether, Absolute Space and Time, \ldots, or Fairies and Witches, was an essential step along
the road to scientific thinking. Probability, too, if regarded as something endowed with
some kind of objective existence, is no less a misleading misconception, an illusory
attempt to exteriorize or materialize our true probabilistic beliefs.
\end{displaycquote}
On the other hand, there may be limits to how certain we can be, no matter how much information we have. Thus, it is common to distinguish between uncertainty that is related to our lack of information and uncertainty that is related to randomness and chance:
\begin{displaycquote}[][italics in original]{o'hagan2004}[]
There are things that I am uncertain about simply because I lack
knowledge, and in principle my uncertainty might be reduced by
gathering more information. Others are subject to random variability,
which is unpredictable no matter how much information I might get;
these are the unknowables. The two kinds of uncertainty have been
debated by philosophers, who have given them the names \textit{epistemic
uncertainty} (due to lack of knowledge) and \textit{aleatory uncertainty} (due
to randomness).
\end{displaycquote}
The word \emph{epistemic} from the Greek \emph{episteme} for knowledge and \emph{aleatory} from the Latin \emph{alea}
for a dice game.

The possibility of aleatory uncertainty appears to challenge \possessivecite{definetti2017} claim.
However, regardless of whether uncertainties originate from a lack of knowledge or from randomness, or whether one believes in randomness at all, the fact is that we remain uncertain:
\begin{displaycquote}[][]{definetti2017}[]
The thing that really matters \ldots{} is the impossible nature of the situation in which we find ourselves when we attempt to foresee a given outcome with certainty. This is so whatever the reason: whether it be ignoreance of certain deterministic laws; or the nonexistence of such laws; or an inability to perform the requisite calculations even though we know the laws; or an inability to obtain precise data
\end{displaycquote}
The probabilities describing our uncertainties always represent our rational degree of belief, even if the cause of our uncertainty is aleatoric.

For example, consider a computer program that outputs either 1 or 0. We do not know the workings of the program, such that the output from the program may be deterministic but unknown to us, such as a computation of the 10,000th digit of pi in binary form, or in some sense random, such as a pseudo-random number generator connected to a source of entropy. These would be epistemic and aleatoric sources of uncertainty, respectively. In any case, if asked to predict the output of the program, we are uncertain.

In the latter case, the entropy source would lead to variation in repeated runs of the program. A particular outcome, 1 say, might occur at a known long-run frequency, $f$. Formally, we can write that frequency as a limit,
\begin{equation}
    f = \lim_{n\to\infty} \frac{k}{n}
\end{equation}
where $k$ is the number of times the program outputs 1 in $n$ runs of the program. \cite{jeffreys1961} and modern treatments~\citep[e.g.,][]{wagenmakers2023} refer to these frequencies as chances. Aleatoric uncertainties are connected to chances.

The connection between epistemic probability and chance was formalized by \possessivecite{definetti2017} representation theorem~\citep[see e.g.,][]{lindley1976}. If we know nothing other than the chance $f$ and if we consider repeat runs of the program to be \emph{exchangeable}, our epistemic probability that the program outputs 1 in the next run must equal the chance $f$. That is, it must equal the long-run frequency at which the program outputs 1,
\begin{equation}
    \text{Pr(Computer outputs 1 in next run)} = f.
\end{equation}
Thus, we can describe randomness using epistemic probability~\citep{spiegelhalter2024}.

In summary, uncertainty is personal and subjective and can be described by epistemic probability, regardless of whether that uncertainty stems from a lack of knowledge or from chance and randomness.

\subsection{Updating}\label{sec:updating}

Having established that probabilities describe uncertainty, we must consider how to update them in light of new information. Here we quickly recapitulate the major results; further details can be found in e.g., \citeand{gregory2005}{sivia2006}.  We use Bayes' theorem,
\begin{equation}
    P(A \given B) = \frac{P(B \given A) \, P(A)}{P(B)},
\end{equation}
for propositions $A$ and $B$. In the context of learning about a model in physics, $M$, using experimental data, $D$, we may write it as,
\begin{equation}\label{eq:posterior}
    P(M \given D) = \frac{P(D \given M) \, P(M)}{P(D)}.
\end{equation}
The terms $P(M)$ and $P(M \given D)$ represent how much we believe in the model before and after seeing the data $D$. The term $P(D \given M)$ is known as the evidence. For a model with unknown parameters $\Theta$, the evidence can be written as an integral,
\begin{equation}\label{eq:evidence}
    P(D \given M) = \int P(D \given M, \Theta) \, p(\Theta \given M) \, \text{d}\Theta
\end{equation}
where $p(\Theta \given M)$ is known as the prior and represents what we knew about the parameters $\Theta$ before seeing any data, and $P(D \given M, \Theta)$ is known as the likelihood. For simplicity, the evidence, likelihood and prior are  often written in shorthand notation as $\mathcal{Z}$, $\mathcal{L}$ and $\pi$, respectively, such that \cref{eq:evidence} may be written as,
\begin{equation}
    \mathcal{Z} = \int \mathcal{L}(\Theta) \, \pi(\Theta) \, \text{d}\Theta.
\end{equation}
Finally, we may wish to compare the several models in light of data. We can do so using the Bayes factor~\citep{jeffreys1961,kass1995},
\begin{equation}
    B_{10} = \frac{P(D \given M_1)}{P(D \given M_0)} .
\end{equation}
This factor tells us how much more we should believe in one model relative to another in light of data;
\begin{equation}\label{eq:posterior_odds}
    \text{Posterior odds} = \text{Bayes factor} \times \text{Prior odds},
\end{equation}
where the prior and posterior odds represent our relative belief in the models before and after seeing the data.

\section{Occam's razor \& fine-tuning}\label{sec:occam}

As well as being uncertain about the weather and capital cities, we are uncertain about scientific models and explanations for experimental data that we have collected. Occam's razor tells us that we should favor simpler models 
\begin{displayquote}[apocryphal]
    entities should not be multiplied beyond necessity
\end{displayquote}
This principle is celebrated in physical sciences and similar ideas were expressed by Aristotle, Ptolemy, Galileo and Newton \citep[for reviews see e.g.][]{sober2015,mcfadden2023}.
In social sciences, however, \cite{gelman2009} expresses skepticism that we should favor models that omit variables or factors for the sake of simplicity. Similar concerns are discussed in machine learning, where big data surely requires big models. Lastly, in biological sciences, \cite{crick1989} warns
\begin{displaycquote}[][]{crick1989}[]
While Occam’s razor is a useful tool
in the physical sciences, it can be a very dangerous implement in biology. It
is thus very rash to use simplicity and elegance as a guide in biological
research \ldots{}

\ldots{} All this may make it very difficult for physicists to adapt to most
biological research. Physicists are all too apt to look for the wrong sorts of
generalizations, to concoct theoretical models that are too neat, too
powerful, and too clean
\end{displaycquote}

\begin{figure}
    \centering
    \includegraphics[height=0.6\linewidth]{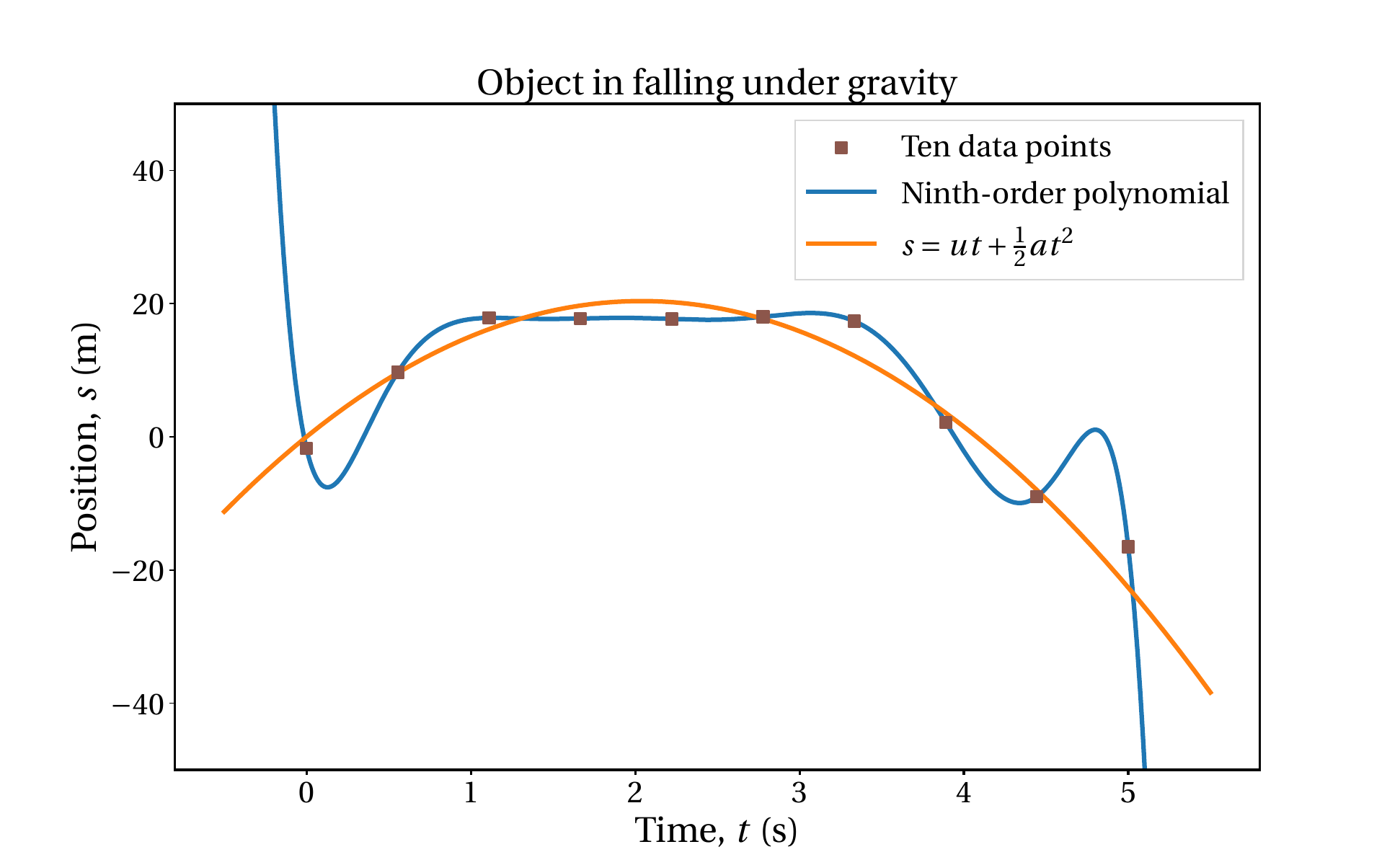}
    \caption{A ninth-order polynomial perfectly fits the ten data points; but should it be preferred over the simpler $s = ut + \frac12 a t^2$?}
    \label{fig:free_fall_10d}
\end{figure}

With these different points of view in mind, we now review Bayesian perspectives on Occam's razor. The modern Bayesian discussion began with \cite{jeffreys1961}, who considered the motion of a ball falling under gravity. According to classical mechanics, the height of the ball $s$ at time $t$ depends on the initial velocity of the ball $u$ and the gravitational acceleration $g$ by
\begin{equation}\label{eq:cm}
    s = u t + \frac12 g t^2.
\end{equation}
On the other hand, more complicated forms such as a polynomial of order $n$
\begin{equation}
    s = \sum_{i=0}^n c_i t^i
\end{equation}
would improve the fit to experimental measurements for $s$ and $t$. For example, see \cref{fig:free_fall_10d}, where ten data points are perfectly fitted by a degree-nine polynomial compared to the imperfect fit from \cref{eq:cm}. \cite{jeffreys1961} felt that simpler \cref{eq:cm} should be preferred and that this was a compelling reason for using Bayesian inference in science,
\begin{displaycquote}[][]{jeffreys1961}[]
    [T]he simplest law is chosen because it is the most likely to give correct predictions; that the choice of based on a reasonable degree of belief; and that the fact that deductive logic provides to explanation of the choice of the simplest law is an absolute proof that deductive logic is grossly inadequate to cover scientific and practical requirements
\end{displaycquote}
Later, \cite{jeffreys1961} formulates Occam's razor as,
\begin{displaycquote}[][italics in original]{jeffreys1961}[]
    \textit{Variation is random until the contrary is shown; and new parameters in laws, when they are suggested, must be tested one at a time unless there is specific reason to the contrary}
\end{displaycquote}
Nevertheless, this discussion is far from satisfactory~\citep{robert2009}. \cite{wrinch1921} implemented Occam's razor manually by assigning priors that favored simpler models in \cref{eq:posterior}; \cite{good1968,good1977} further developed this approach.

Stimulated by attempts to formulate a Bayesian account of science~\citep{rosenkrantz1977}, \cite{jaynes1979} considered Occam's razor in the context of Bayesian inference:
\begin{displaycquote}[][]{jaynes1979}[]
Ockham says that we should prefer
the simpler [hypothesis] \ldots{} intuition assents at once. But this only
set the stage for centuries of discussion over precisely what is meant
by simplicity \ldots{} It is interesting to see the mechanism by which
Bayes’s theorem usually justifies but in some cases modifies this
intuition
\end{displaycquote}
By computing the evidences for a simple model and a model that extended it by adding a new parameter, \cite{jaynes1979} found that
\begin{displaycquote}[][]{jaynes1979}[]
 if the old model already is flexible enough to accommodate
the data, then, as a general rule, Bayes' theorem will, like Ockham,
tell us to prefer the intuitively simpler hypothesis \ldots

But, having seen this mechanism, it is easy to invent cases \ldots{} in which Bayes’s theorem will contradict Ockham because it is taking into account further circumstances undreamt of in
Ockham’s philosophy \ldots{} Our conclusions depend crucially on the prior information \ldots{} Even in Bayesian theory, the
question is subtle enough to have caused trouble.
\end{displaycquote}
Shortly after \cite{jaynes1979} and independently, \cite{smith1980} reached a similar conclusion.
\begin{displaycquote}[][]{smith1980}[]
[T]he
Bayes factor is seen to function as a fully automatic Occam's Razor-cutting back to the simpler
model whenever there is nothing to be lost by so doing
\end{displaycquote}
Unlike that considered by \cite{wrinch1921}, \citeand{jaynes1979}{smith1980} describe an \emph{automatic} form of Occam's razor captured inside Bayesian inference. 

The question of Occam's razor was again considered in the late 1980s and early 1990s in the maximum entropy conference series~\citep{gull1988,loredo1990,garrett1991a,garrett1991b}. To quantify the automatic razor, \cite{gull1988} decomposed the evidence into goodness-of-fit and an Occam term,
\begin{equation}\label{eq:z_decomp_fit_occam}
    \mathcal{Z} = \textstyle\max_{\,\Theta} \mathcal{L} \times \mathcal{O}
\end{equation}
where, by definition,
\begin{equation}
     \mathcal{O} \equiv \frac{\mathcal{Z}}{\max_{\,\Theta} \mathcal{L}}
\end{equation}
Thus, the Bayes factor factorizes into a likelihood ratio and an \emph{Occam factor},
\begin{equation}
    B_{10} = \frac{\mathcal{Z}_1}{\mathcal{Z}_0} = \frac{\max_{\,\Theta_1} \mathcal{L}_1}{\max_{\,\Theta_0} \mathcal{L}_0} \times \frac{\mathcal{O}_1}{\mathcal{O}_0}
\end{equation}
In other words, the Occam factor was the automatic penalty that arises in Bayesian statistics from \emph{averaging} rather than \emph{maximizing} over a model's unknown parameters. The decomposition in \cref{eq:z_decomp_fit_occam} may be generalized as
\begin{equation}
    \log \mathcal{Z}  = \langle \log \mathcal{L} \rangle_{\mathcal{Q}} - \langle \log {\mathcal{L}}/{\mathcal{Z}} \rangle_{\mathcal{Q}} 
\end{equation}
for any choice of distribution $\mathcal{Q}$ for the model parameters, where $\langle \cdot \rangle_{\mathcal{Q}} = \int \cdot q(\Theta) \,\text{d}\Theta$ denotes an expectation. Taking $\mathcal{Q}$ as a Dirac mass at the best-fit parameters yields \cref{eq:z_decomp_fit_occam}. Taking $\mathcal{Q}$ as the posterior distribution  $\mathcal{P}$ yields,
\begin{equation}
    \ln \mathcal{Z}  = \langle \log \mathcal{L} \rangle_{\mathcal{P}} - H,
\end{equation}
where $H$ is the Kullback-Liebler divergence between the prior and posterior. This was recently advocated by \cite{hergt2021} and supplies an information-theoretic meaning to the Occam term.

\begin{figure}
    \centering
    \includegraphics[height=0.6\linewidth]{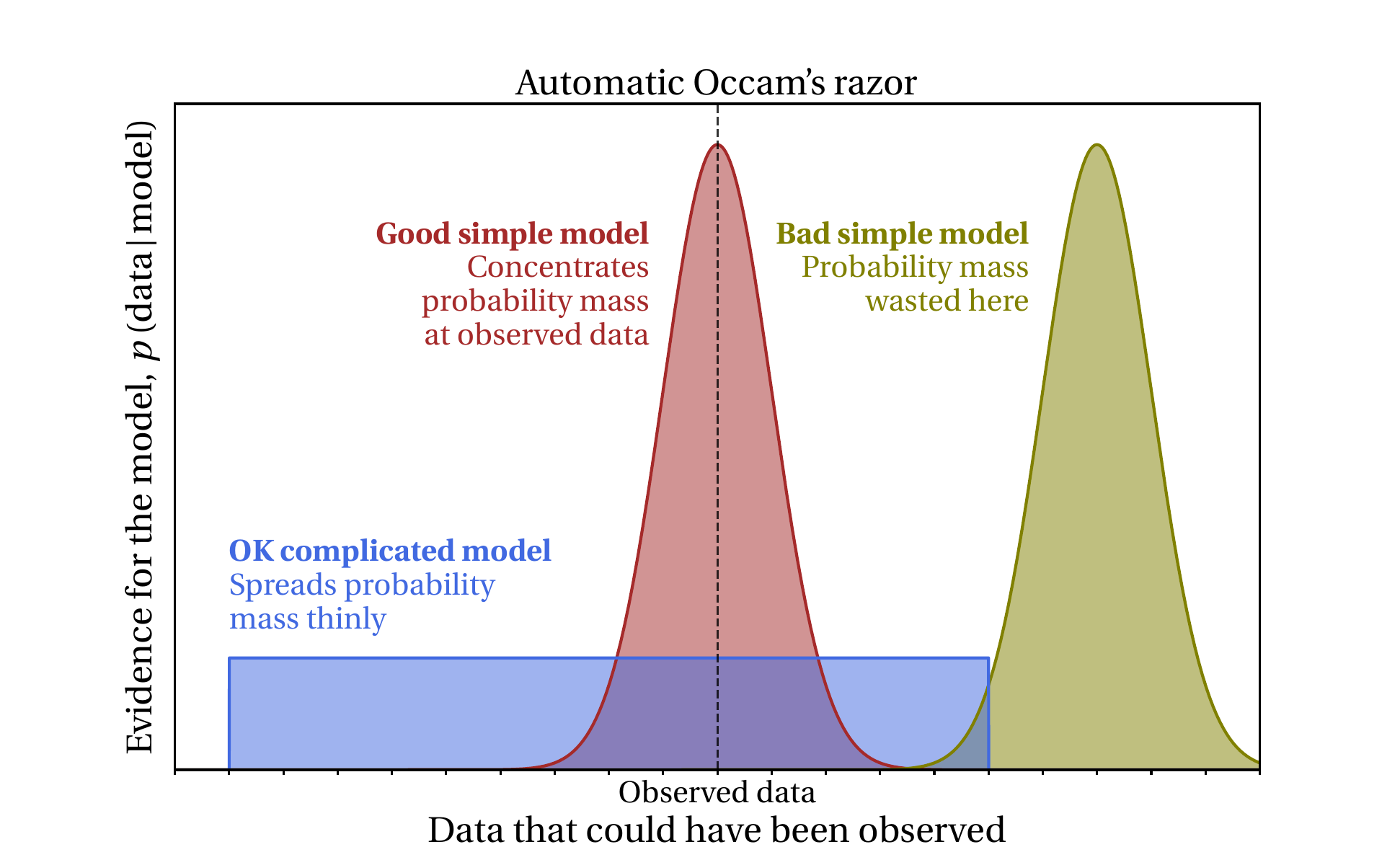}
    \caption{\possessivemulticite{mackay1992a}{mackay1992b} explanation of the automatic Occam's razor: complicated models spread their predictions thinly.}
    \label{fig:mackay_occam}
\end{figure}

\possessivecite{jeffreys1961} example of a ball in free-fall was picked up by \cite{jefferys1991,jefferys1992a,jefferys1992b}, who concluded:
\begin{displaycquote}[][]{jefferys1992a}[]
Ockham's razor, far from being merely an ad hoc principle, can under many practical
situations in science can be justified as a consequence of Bayesian inference
\end{displaycquote}
\cite{mackay1991,mackay1992a,mackay1992b,mackay1992c,mackay1992d,mackay1992e,mackay2003} continued discussion of an automatic Occam's razor in the context of machine learning, neural networks and regression:
\begin{displaycquote}[][]{mackay1992b}[]
The problem of Occam’s razor rears its head repeatedly when we try to make
these design choices because a more complex and unconstrained neural network will nearly
always learn the examples in the training set better than a simpler one; however the simpler
neural network may actually be a better model of the problem, and generalise better to new
examples
\end{displaycquote}
\cite{mackay1991,mackay1992a,mackay1992b,mackay1992c,mackay1992d,mackay1992e,mackay2003} introduces \cref{fig:mackay_occam} to explain the automatic razor.  Simple models make sharp predictions for the observed data; complicated models make broad predictions. Considering the evidence as a distribution in data, complicated models dilute the evidence by making broad predictions. \cite{murray2005} further discussed this illustration of the automatic razor through examples. The discussion of Occam's razor in neutral network architecture continued in \cite{rasmussen2000};
\begin{displaycquote}[][]{rasmussen2000}[]
One might think that one has to build a prior
over models which explicitly favours simpler models. But as we will see, Occam's Razor is
in fact embodied in the application of Bayesian theory.
\end{displaycquote}
In this setting of big data, we anticipate that complexity should be reflected in neutral network architecture. 
\cite{rasmussen2000} discussed whether architecture should be determined by an automatic Occam's razor and the notion of Occam's hill --- the peak tradeoff between fit and complexity. See \cite{lotfi2022} for more recent discussions of Occam's razor in machine learning. Later, \cite{balasubramanian1997} presented the automatic razor through analogies with statistical mechanics.

By the mid-nineties, the automatic razor was well-known and mentioned in the seminal review of Bayes factors by \cite{kass1995}. The existence of an automatic razor is thus now well established. This leaves, however, the question of whether one should add a manual razor and explicitly penalize models by a measure of complexity. We believe that the situation is best summarized by 
\begin{displaycquote}[][]{neal1996}[]
[W]e might ask whether Occam's Razor is of any use to
Bayesians \ldots{} Bayesians needn't concern themselves with Occam's
Razor, since to the extent that it is valid, it will be applied automatically anyway.
\end{displaycquote}

\subsection{\possessivecite{jeffreys1961} example}

Armed with a modern understanding of the automatic Occam's razor, we return to \possessivecite{jeffreys1961} example of the quadratic rule for an object's height under free-fall \cref{eq:cm}. For simplicity, we compare a quadratic and a cubic law,
\begin{align}
    s &= u t + \frac12 a t^2\\
    s &= u t + \frac12 a t^2 + c t^3
\end{align}
with unknown parameters $u$, $a$ and $c$. We consider four measurements $(t_i, s_i)$, since, as we shall see, this is all we need to make our point.

To use illustrate the automatic razor in action, we decompose the evidences into factors for each measurement,
\begin{equation}
    P(s_1, s_2, s_3, s_4 \given M) = P(s_1 \given M) \, P(s_2 \given s_1, M) \, P(s_3 \given s_1, s_2, M) \, P(s_4 \given s_1, s_2, s_3, M)
\end{equation}
Thus, the Bayes factor may be written as product of partial Bayes factors,
\begin{align}\label{eq:partial_bayes_factors}
    B_{10} ={}& \frac{P(s_1, s_2, s_3, s_4 \given M_1)}{P(s_1, s_2, s_3, s_4 \given M_0)}\\
           ={}&\frac{P(s_1 \given M_1)}{P(s_1 \given M_0)} \frac{P(s_2 \given s_1, M_1)}{P(s_2 \given s_1, M_0)} \frac{P(s_3 \given s_1, s_2, M_1)}{P(s_3 \given s_1, s_2, M_0)} \frac{P(s_4 \given s_1, s_2, s_3, M_1)}{P(s_4 \given s_1, s_2, s_3, M_0)}\\
           ={}&  P_1 \times P_{2 \given 1} \times P_{3 \given 1, 2} \times P_{4 \given 1,2,3}
\end{align}
where for shorthand e.g. 
\begin{equation}
    P_{2 \given 1} \equiv \frac{ P(s_2 \given s_1, M_1)}{ P(s_2 \given s_1, M_0)}
\end{equation}
denotes the partial Bayes factor.

\begin{figure}
    \centering
    \includegraphics[width=0.97\linewidth]{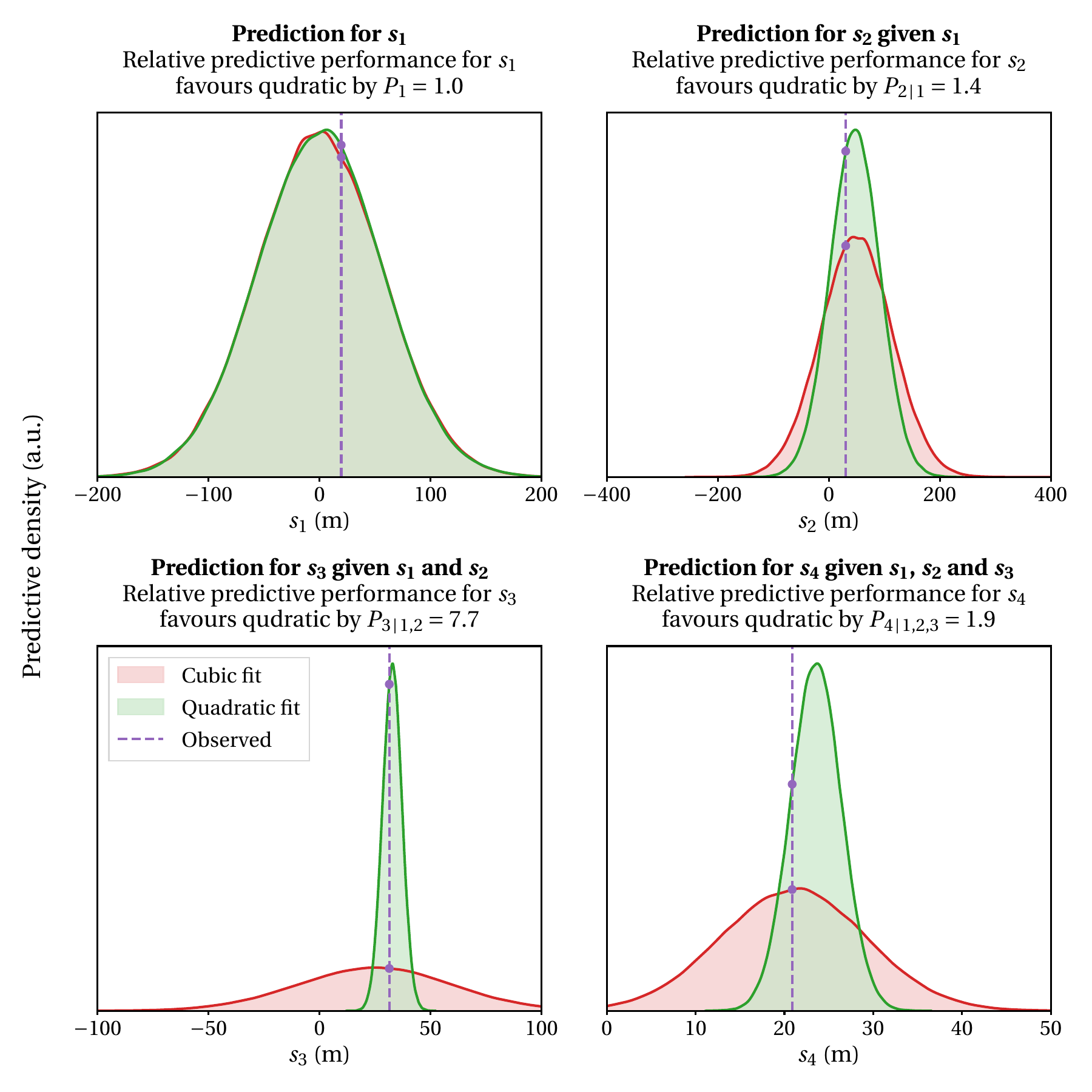}
    \caption{Partial Bayes factors, \cref{eq:partial_bayes_factors}, for four data points comparing quadratic and cubic laws for an object's trajectory.}
    \label{fig:free_fall_mackay_occam}
\end{figure}

In \cref{fig:free_fall_mackay_occam} we plot these factors as distributions in the data in analogy to \cref{fig:mackay_occam}. We see that for the first data point $s_1$ the quadratic law outpredicts the cubic law by $1.2$. That one data point doesn't tell us much, and the predictions from both models for $s_2$ remain broad, though the quadratic outpredicts the cubic law by $2.9$. With two data points, the quadratic law's two parameters are somewhat determined, and it makes a sharp prediction for $s_3$. The cubic law's prediction, on the other hand, remains broad, since that laws three parameters remain underconstrained. The quadratic model thus outpredicts the cubic model by about $18$. Lastly, with three data points fitted, both models make somewhat specific predictions for the remaining point $s_4$, with the quadratic law outpredicting the cubic by about $2$. Taking the product of these partial Bayes factors, we find that the four measurements favor the quadratic law by about $120$.

\begin{figure}
    \centering
    \includegraphics[width=0.97\linewidth]{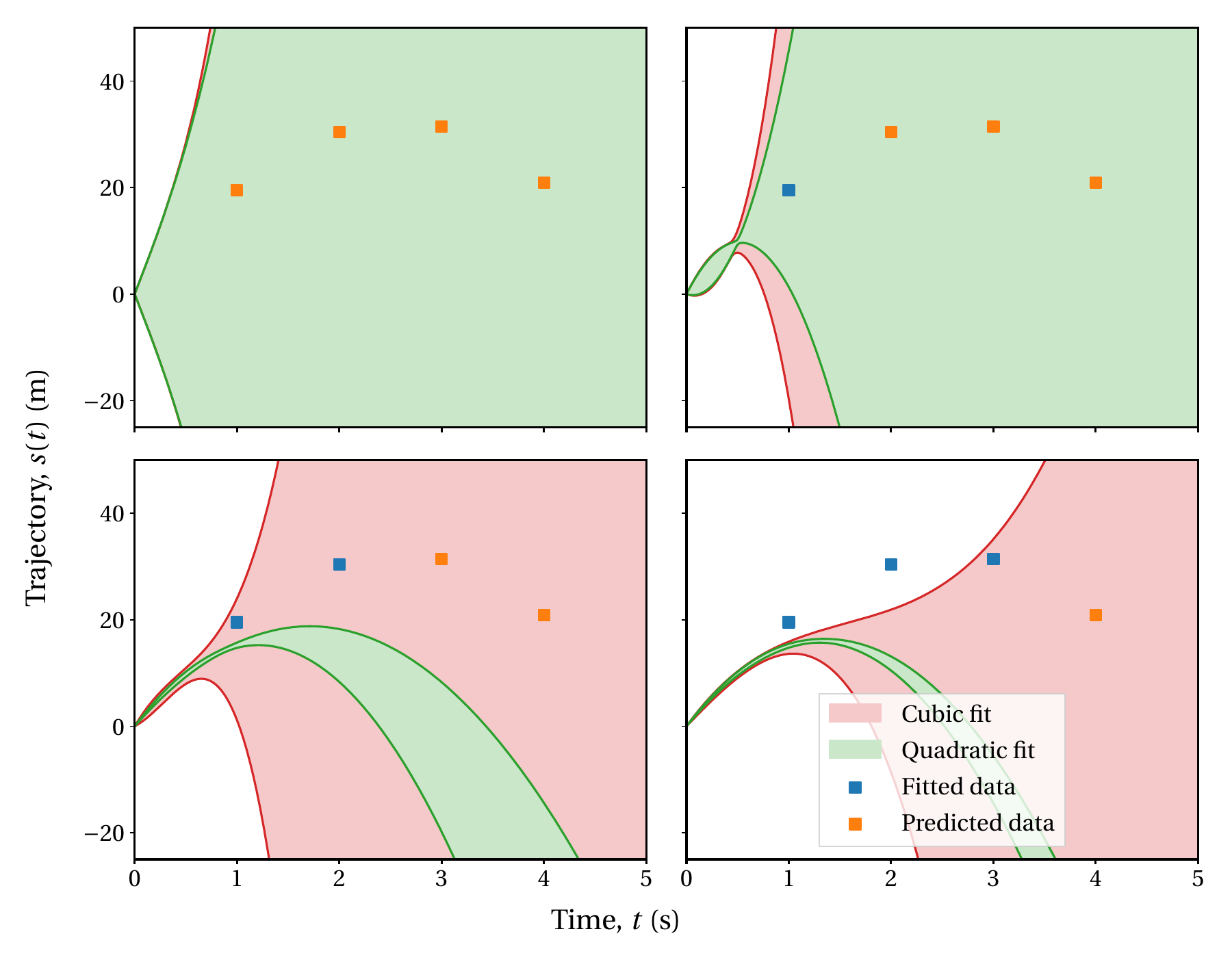}
    \caption{Uncertainty in trajectory under quadratic and cubic laws.}
    \label{fig:free_fall_trajectory}
\end{figure}

As \cite{jaynes1979} warned, however, the extent to which the Bayes factor contains an automatic razor depends on our prior information. In \cref{fig:free_fall_mackay_occam} the prior for our cubic coefficient was a normal centered at zero with width $\delta = 50\,\text{m/s}^3$. Was that appropriate? We need to think about things, not Greek letters~\citep{lindley2006}. What effect were we modeling with a cubic term? What do we know about that effect? This cubic term corresponds to a time-dependent acceleration, $a(t) = a + c t$. 

\begin{figure}
    \centering
    \includegraphics[height=0.6\linewidth]{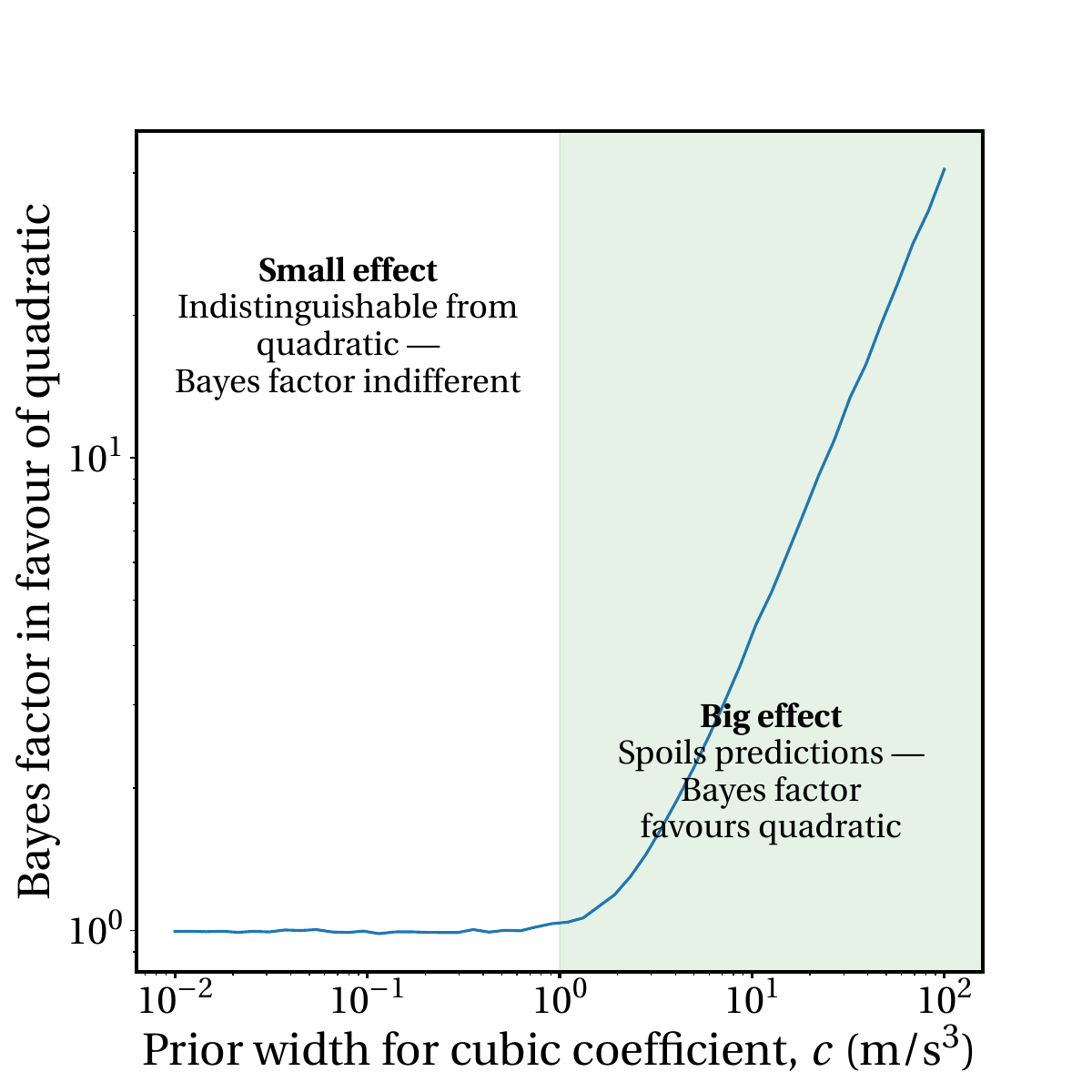}
    \caption{Prior sensitivity of Bayes factor in favor of quadratic versus cubic law.}
    \label{fig:free_fall_prior_sensitivity}
\end{figure}

If we were modeling possible time-dependence in the gravitational constant, it might be reasonable to consider it as background knowledge that time-dependence must be small, $\delta \ll 1\,\text{m/s}^3$. On the other hand, suppose the object was a sheet of paper. If we were modeling the fluttering of the paper as it fell downwards, we might anticipate a moderate effect on the paper's downward acceleration, $\delta \gg 1\,\text{m/s}^3$. In \cref{fig:free_fall_prior_sensitivity} we show the Bayes factor as a function of the choice $\delta$. If $\delta \gg 1\,\text{m/s}^3$, the Bayes factor punishes the cubic model for making broad predictions, as in \cref{fig:free_fall_mackay_occam}. On the other hand, if $\delta \ll1\,\text{m/s}^3$, the Bayes factor is indifferent between the quadratic and cubic. The latter result contradicts Occam's razor, which would disfavour the cubic model.

In the latter case, which model should we pick, if any? It's worth considering the prior odds at this point. If we strongly believed that the gravitational constant was time-dependent, we should favor the cubic law in our assigned prior odds. We should expect, furthermore, that the cubic better predicts future data. There is no harm in including effects that are irrelevant for explaining our current data; they may be relevant for predicting future data.

\section{Naturalness}\label{sec:naturalness}

Although the literature on naturalness and fine-tuning is vast, beginning at least as early as  \citeand{Wilson:1970ag,Susskind:1978ms}{tHooft:1979rat}, discussions about the connections between epistemic probability and naturalness are somewhat limited. The connections between probability and traditional measures of naturalness~\citep{Barbieri:1987fn} were explored informally by \citeand{deCarlos:1993rbr,anderson1995,Ciafaloni:1996zh,strumia1999}{athron2007}. An explicitly Bayesian approach to naturalness was first applied to parameter inference in supersymmetric models through a choice of priors. \citeand{giusti1999}{allanach2006} constructed priors that explicitly disfavored fine-tuning; whereas \citeand{allanach2007}{cabrera2009,cabrera2010} later identified that an automatic penalty was present, if one parametrised a model in terms of fundamental parameters. Later, \cite{fichet2012} demonstrated that the Bayes factor was most relevant in this setting. This was expanded on by \citeand{balazs2013,kim2014,fowlie2014b,fowlie2014a,fowlie2016,athron2017,clarke2017}{fundira2018}. The issue was further explored in theses~\citep{fowlie2013,farmer2015,murnane2019,koren2020}. Lastly, traditional fine-tuning measures were identified exactly with a so-called Bayes factor surface~\citep{fowlie2024,fowlie2025}.

Despite this activity, the words Bayes and variants are completely absent from major reviews of naturalness in physics~(e.g.,~\cite{craig2023,peskin2025}), though they are occasionally discussed in philosophy~\citep{Williams:2015gxa,grinbaum2012,williams2018,wallace2019,bain2019,borrelli2019,fischer2023,fischer2024}. We believe \cite{williams2018} sheds light on this: there are (at least) two  interpretations of naturalness. First, it is connected to improbable cancellations, formalized by the use of probability theory. Second, it concerns the autonomy of scales,
\begin{displaycquote}[][]{williams2018}[]
    [The] statistical notion of naturalness has become widespread,
leading to a bifurcation of naturalness into two notions which are closely related,
both historically and conceptually, but essentially distinct: one notion of naturalness
according to which naturalness problems are failures of an expectation about the
autonomy of scales, and a second notion according to which naturalness problems stem
from a parameter (or theory) being ``unlikely'' or ``improbable''
\end{displaycquote}
The mixing between infra-red and ultra-violet scales caused by quadratic corrections violates this autonomy. The desire to maintain an autonomy of scales is often justified by a retelling of historical predictions and discoveries, e.g.~\possessivecite{weisskopf1939} postdiction of the positron and \possessivecite{Gaillard:1974hs} prediction of the charm quark. 

Unfortunately, as a consequence of this bifurcation, the statistical foundations of naturalness are poorly understand and naturalness is considered by some to be a matter of taste, e.g.,
\begin{displaycquote}[][]{Shifman:2012na}[]
    The criterion of
naturalness is aesthetic, or, if you wish, philosophic. If you do not like it you can
ignore it. Most people like it
\end{displaycquote}
Thus we now recapitulate how the automatic Occcam's razor favors natural theories.

\subsection{Bayesian naturalness}

As an example of a naturalness argument, we consider the hierarchy problem~\citep{peskin2025}, though we anticipate that our reasoning applies to other naturalness problems. We make things as simple as possible to demonstrate the impact of quadratic corrections to the weak scale. We represent the weak scale by the $Z$ boson mass, which was measured to be about $\hat M_Z \simeq 100\,\text{GeV}$, and take the Planck scale as $\Lambda = 10^{18}\,\text{GeV}$.

First, we consider a model without quadratic corrections that predicts,
\begin{equation}
    M_0: \quad M_Z^2 = \mu^2.
\end{equation}
The predicted $Z$ boson mass is equal to the unknown parameter of the model, $\mu$. Second, we consider a model in which there are Planck-scale quadratic corrections to the predicted $Z$ boson mass,
\begin{equation}\label{eq:z_mass_with_quadratic_corrections}
    M_1: \quad M_Z^2 = -\mu^2 + \Lambda^2.
\end{equation}
where again $\mu$ is an unknown parameter. For simplicity, in this case we consider a positive quadratic correction and a $\mu^2$-parameter that appears with a negative sign.

In the first case, the predicted distribution for the $Z$ boson mass equals the prior distribution for the $\mu$-parameter;
\begin{equation}
    p(\log M_Z \given M_0) = \pi(\log \mu = \log M_Z \given M_0).
\end{equation}
In the second case, there is a Jacobian factor,
\begin{align}
    p(\log M_Z \given M_1) =& \left|\frac{d \log \mu}{d \log M_Z}\right| \, \pi(\log \mu = \log \sqrt{\Lambda^2 - M_Z^2} \given M_1)\\
                         =& \left|\frac{M_Z^2}{\Lambda^2 - M_Z^2}\right| \, \pi(\log \mu = \log \sqrt{\Lambda^2 - M_Z^2} \given M_1)
\end{align}
Thus, Bayes factor,
\begin{align}\label{eq:bf_general}
    B_{10} =& \frac{p(\log M_Z \given M_1)}{p(\log M_Z \given M_0)} \\
    =&\left|\frac{M_Z^2}{\Lambda^2 - M_Z^2}\right| \, \frac{\pi(\log \mu = \log \sqrt{\Lambda^2 - M_Z^2} \given M_1)}{\pi(\log \mu = \log M_Z \given M_0)}
\end{align}
The factor ${M_Z^2} / {(\Lambda^2 - M_Z^2)}$ originates from the relationship \cref{eq:z_mass_with_quadratic_corrections} between $\log M_Z$ and $\log \mu$, rather than from any prior choices.
Because $\hat M_Z \lll \Lambda$, we anticipate that $\hat M_Z^2 / (\Lambda^2 - M_Z^2) \simeq \hat M_Z^2 / \Lambda^2 \lll 1$ and thus $B_{10} \lll 1$ for the observed $Z$ boson mass. We should, however, check the impact of the ratio of priors. 

\subsection{Scale-invariant priors}

\begin{figure}
    \centering
    \includegraphics[width=0.97\linewidth]{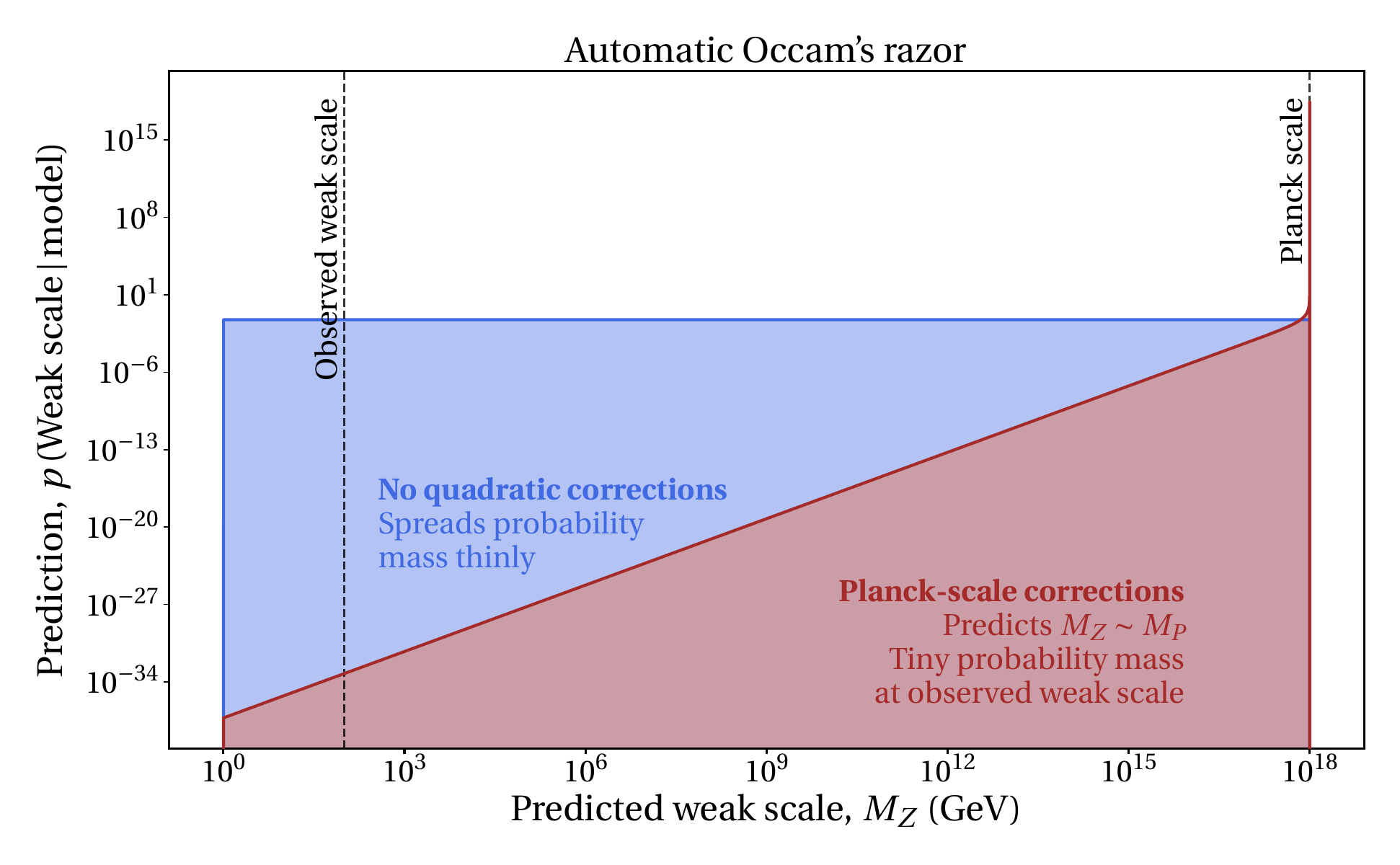}
    \caption{Predictions for the weak scale with (red) and without (blue) Planck-scale quadratic corrections for $p(\ln \mu) = \text{const}$. The predictive density at the observed weak scale disfavors the model with quadratic corrections by about $10^{32}$.}
    \label{fig:hierarchy_mackay_occam}
\end{figure}

First, suppose that we consider our prior knowledge of $\mu$ in both models equivalent. In that case, we should assign an identical prior to the $\mu$-parameter in each case,
\begin{equation}
\pi(\log \mu \given M_1) = \pi(\log \mu\given M_0)
\end{equation}
Suppose we are ignorant of the scale of the $\mu$-parameter. In this case, we should choose
\begin{equation}
\pi(\log \mu \given M_1) = \pi(\log \mu\given M_0) = \text{const.}
\end{equation}
This is a scale-invariant or logarithmic prior. We show predictions for the $Z$ mass in \cref{fig:hierarchy_mackay_occam} with and without quadratic corrections.
For any observed $Z$ mass, the Bayes factor equals,
\begin{equation}
     B_{10} = \frac{M_Z^2}{\Lambda^2 - M_Z^2}
\end{equation}
For the observed $Z$ boson mass, this leads to $B_{10} \simeq 10^{-32}$. 
In other words, the hierarchy $\Lambda \ggg \hat M_Z$ leads to overwhelming preference for the model without quadratic corrections. 

To avoid $B_{10} \lll 1$, we could broaden the prior for the $\mu$-parameter only in the model without quadratic corrections. We don't know why conditioning on quadratic corrections would change the plausible range for the $\mu$-parameter, but pursue this choice for the sake of argument.  Considering scale-invariant priors for the $\mu$-parameter, but between $(a_0, b_0)$ and $(a_1, b_1)$ in the models 0 and 1, respectively, we find
\begin{equation}\label{eq:bf_log_priors_different_ranges}
     B_{10} = \frac{\hat M_Z^2}{\Lambda^2 - \hat M_Z^2} \, \frac{\log b_0 / a_0}{\log b_1 / a_1}  
\end{equation}
If we take,
\begin{equation}
    \log b_0 / a_0 = \frac{\Lambda^2 - \hat M_Z^2}{\hat M_Z^2} \log b_1 / a_1 
\end{equation}
we can cancel the first factor in \cref{eq:bf_log_priors_different_ranges} so that $B_{10} = 1$. This would mean taking the $\mu$-parameter between to lie between $n$ orders of magnitude in the presence of quadratic corrections, but between about $n \times 10^{32}$ orders of magnitude without quadratic corrections. This path around the hierarchy problem thus seems contrived.

\subsection{Exploring other choices}

\begin{figure}
    \centering
    \includegraphics[width=0.97\linewidth]{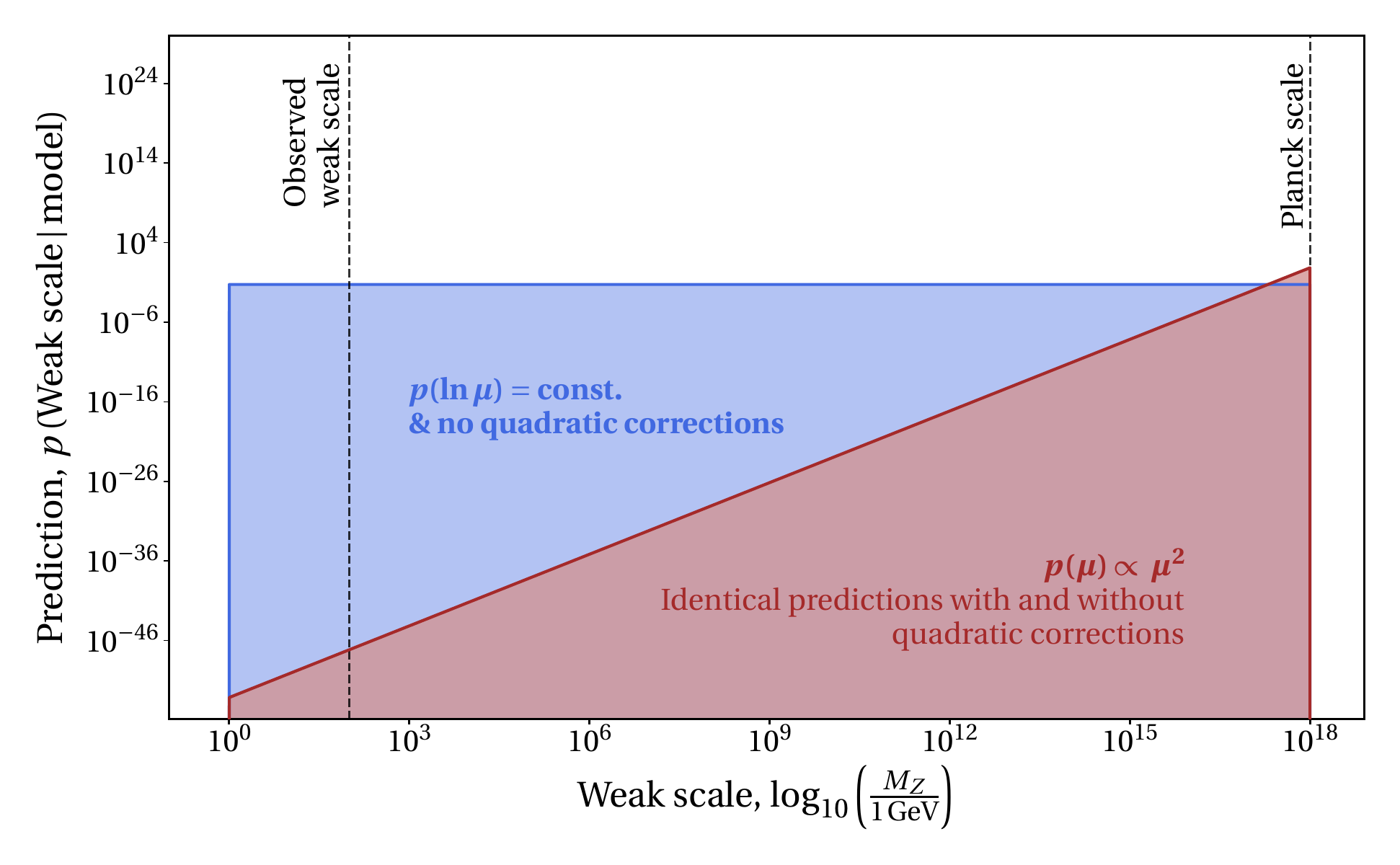}
    \caption{Predictions for the weak scale with no quadratic corrections and $p(\ln \mu) = \text{const}$ (blue), and $p(\mu) \propto \mu^2 $ (red).}
    \label{fig:hierarchy_bf_1}
\end{figure}

From \cref{eq:bf_general}, we see that we can make $B_{10} \ge 1$ for the observed $Z$ boson mass if
\begin{equation}
    \pi(\log \mu \given M_1) = \pi(\log \mu\given M_0) \propto \mu^{2 + n} \quad \text{for } n \ge 0
\end{equation}
since this leads to
\begin{equation}
     B_{10} = \left|\frac{\Lambda^2 - M_Z^2}{M_Z^2}\right|^n
\end{equation}
and $\Lambda \ggg \hat M_Z$. This choice thus appears to ameliorate the hierarchy problem. On the other hand, consider the prior predictive distribution for the $Z$ mass for such a choice in \cref{fig:hierarchy_bf_1}. We see that with this prior we predict $M_Z \sim \Lambda$ in both models; the Bayes factor was altered by making both models equally bad at predicting the $Z$ boson mass. 

\begin{figure}
    \centering
    \includegraphics[width=0.97\linewidth]{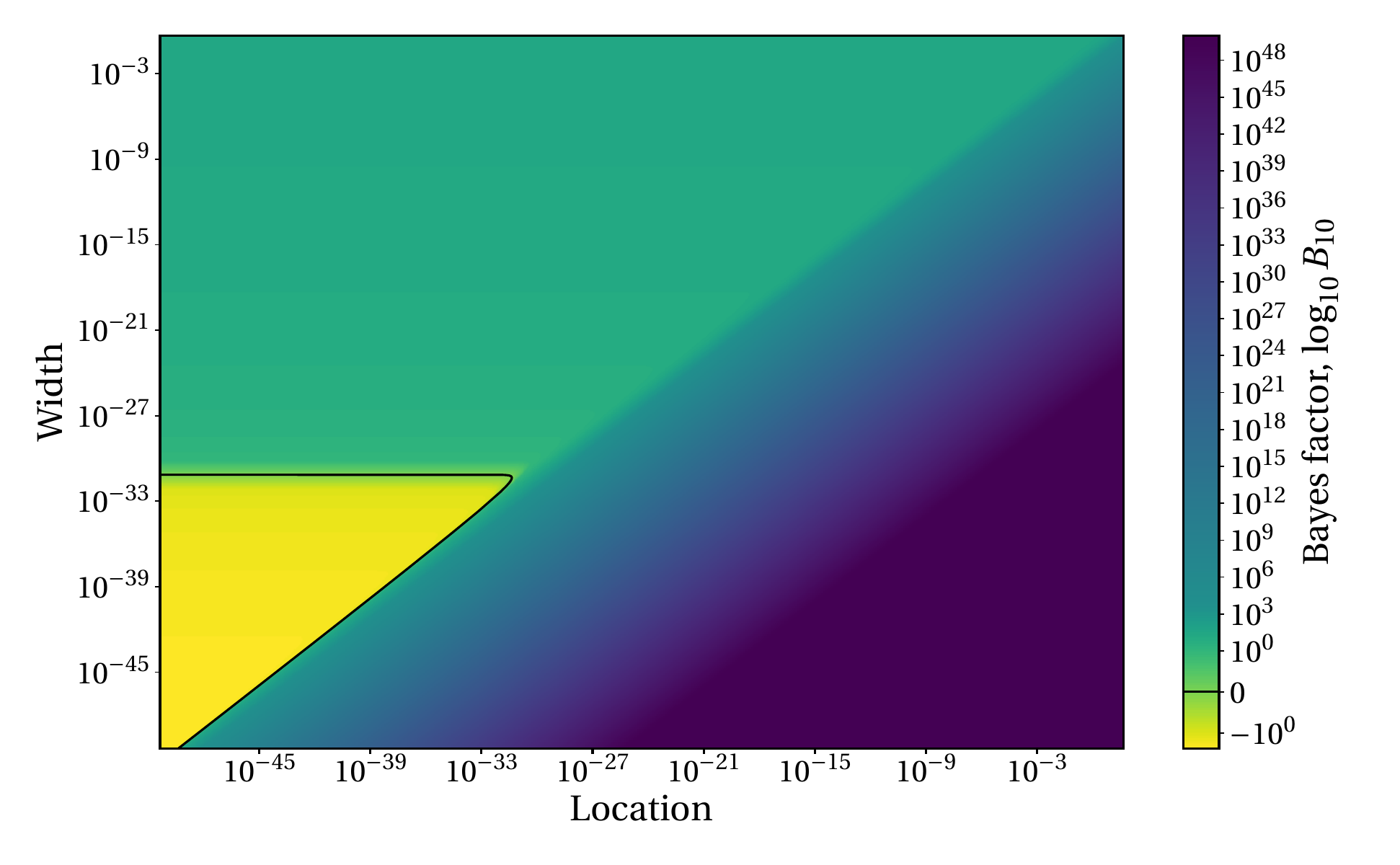}
    \caption{Bayes factor in favour of model without quadratic corrections. In presence of quadratic corrections, the prior for the $\mu$-parameter centered around $\mu_Z$ to achieve $M_Z \lll M_P$;  $\log \mu  \sim \mathcal{N}(\log \mu_Z - d, w^2)$. }
    \label{fig:prior_pred_mz_normal}
\end{figure}

Finally, let us consider only changing the prior for the $\mu$-parameter in the model with quadratic corrections. From \cref{eq:bf_general} we see that we need to construct a prior that peaks at 
\begin{equation}
    \log \mu \simeq \log \sqrt{\Lambda^2 - \hat M_Z^2} \equiv \log \mu_Z
\end{equation}
We consider a normal distribution for $\log\mu$,
\begin{equation}
    \log \mu  \sim \mathcal{N}(\log \mu_Z - d, w^2),
\end{equation}
peaked at $d$ away from $\log \mu_Z$ and with width $w$.
The resulting Bayes factor is shown in \cref{fig:prior_pred_mz_normal} as a function of the location $d$ and width $w$ hyperparameters. We see that the prior must be extraordinarily narrow and focused at $\log\mu_Z$ to achieve $B_{10} > 1$.

\section{Naturalness arguments are all in your head}\label{sec:discuss}

We now finally return to our original goal: clearing up misunderstandings in recent evaluations of naturalness~\citep{hossenfelder2019,wells2025}. \cite{wells2025} claims that aleatoric uncertainty in a theory's parameters is a premise of all naturalness arguments. Specifically, the parameters must originate from randomness:
\begin{displaycquote}[][]{wells2025}[]

    Coefficients of the operators of the Ur-Theory are aleatorily
assigned to each of the symmetry-allowed operators \ldots

What that means is that they are randomly selected according to some principles
and setting that is not entirely known to us, but the parameters are nevertheless contingent.
They might arise from Wotan throwing dice or from quantum mechanical fluctuations \ldots{} The premise is agnostic to the precise
mechanism, but it insists that the parameters of the theory are selected somehow, and that
it has an element of contingency.
\end{displaycquote}
Regardless of the source of uncertainty, however, we are uncertain about a model's parameters. The parameters are unknown to us. There is no need for randomness, whatever it may mean. \cite{wells2025} elaborates that
\begin{displaycquote}[][]{wells2025}[]
[I]t all
becomes a probability question. Without a probability density existence supposition of one
kind or another the exercise is meaningless \ldots

[T]here is no Hierarchy Problem if there is no contingency. We are the one
and only universe, not born in a distribution of many possibilities. The parameters were not
determined in a dice game between Wotan and his friends. They are just so. And if they are
just so, there can be no discussion of improbable. Declaring a special value to be improbable
in such circumstances would be committing a fallacy of illicit probabilistic inference.
\end{displaycquote}
The use of probability to describe epistemic uncertainty lies at the heart of Bayesian inference; it is perplexing to call it an illicit or fallacious use of probabilistic inference.

In a similar vein, \cite{hossenfelder2019} argues against the use of probability theory in the context of naturalness:
\begin{displaycquote}[][]{hossenfelder2019}[]
A probability distribution from which to calculate the most likely choice of parameter adds
unnecessary structure to the theory and is thus in conflict with the dictum of simplicity. We
could have chosen a parameter and be done with it. The probability distribution and all the
not-observed values of the parameters are unnecessary for the derivation of any observable
and they should therefore be stripped by Occam’s razor.
\end{displaycquote}
This is based on a misunderstanding about the nature of probability in this setting. The parameters are unknown to us. The probabilities represent our uncertainty about them; the probabilities are epistemic. The probabilities are not objective; they are not part of the theory or the theory's ontology. Thus, they cannot be said to be adding structure to the theory.

\cite{hossenfelder2019} continues,
\begin{displaycquote}[][]{hossenfelder2019}[]
No one in their right mind would start with defining a useless
probability distribution over a space from which eventually only one value is needed.
\end{displaycquote}
We do not know if anyone would \emph{start} that way, though prior probability distributions were used in physics by e.g., Planck and LIGO, and of course throughout any field that uses Bayesian inference. One value won't do; we need a distribution to describe our uncertainty, because we don't and can't ever know with certainty the exact value of a parameter.

\cite{hossenfelder2019} ignores the automatic razor in Bayesian inference:
\begin{displaycquote}[][]{hossenfelder2019}[]
The Bayesian approach to technical naturalness is merely a different way to quantify the sensitivity of the low-energy parameters on the high-energy
parameters. This is a good way to avoid having to pick one particular measure for
naturalness. But I don’t question the sensitivity itself; I question it is rational to believe a theory less sensitive to high energies is more likely to be correct. The Bayesian
approach doesn’t say anything about this.
\end{displaycquote}
The Bayesian approach is loud and unmistakable here, as shown in \cref{sec:naturalness}. \cite{hossenfelder2019} continues by suggesting that the prior odds in \cref{eq:posterior_odds} were neglected:
\begin{displaycquote}[][]{hossenfelder2019}[]
These Bayesian assessments, however, do not quantify the presence of the additional
axiom which is the symmetry itself (or whatever other assumption it is that makes
a model natural) \ldots

\ldots{} there is no reason to assume that a theory is more likely to be a better description of
nature just because it is more rigid.
\end{displaycquote}
This is true: one could have reason to consider theories without quadratic corrections to be about $10^{32}$ times less plausible a priori than theories with quadratic corrections. In this case, the prior odds would scotch the factor of $10^{32}$ coming from the fact the hierarchy between the weak and Planck scales. Even in this case, though, the fact that the hierarchy \emph{changed} our belief by factor of $10^{32}$ remains remarkable.

In amongst this discussion, \cite{hossenfelder2019} remarks that Bayesian model comparison may be common, but it might be bad practice:
\begin{displaycquote}[][]{hossenfelder2019}[]
It is common practice in the literature of Bayesian assessments to
compare models with different assumptions (rather than just the same model with
different parameters), but that doesn't mean it's good practice
\end{displaycquote}
There are indeed critics of Bayesian model comparison~\citep[e.g.,][]{gelman1995} and specifically Bayes factors~\citep[e.g.,][]{cousins2008,robert2015}, and there is scope for their misuse~\citep{tendeiro2024}. On the other hand, more specific arguments are required here.

In summary, the connections between Bayesian inference and fine-tuning have been explored in various fields starting from the late 1970s, and reaching naturalness and high-energy physics in the 2000s. Historically, however, naturalness was commonly interpreted as a requirement for autonomy of scales, such that the foundations for naturalness arguments offered by Bayesian inference were overlooked. As a consequence, there remains confusion about the epistemic nature of probability in naturalness arguments. In particular, the probabilities in Bayesian treatments of naturalness were wrongly interpreted as aleatoric in the recent literature. These probabilities are in fact epistemic --- all in your head --- and do not add to a theory's ontology or make assumptions about a physical mechanism for randomness.

\subsection*{Acknowledgements}

I was supported by RDF-22-02-079 and NSFC RFIS-II W2432006.
I would like thank James Wells for clarifications.

\bibliographystyle{custom}
\bibliography{refs}

\begin{thebibliography}{96}
\providecommand{\natexlab}[1]{#1}
\providecommand{\url}[1]{\texttt{#1}}
\expandafter\ifx\csname urlstyle\endcsname\relax
  \providecommand{\doi}[1]{doi: #1}\else
  \providecommand{\doi}{doi: \begingroup \urlstyle{rm}\Url}\fi

\bibitem[Allanach(2006)]{allanach2006}
 Allanach~B.~C., (2006). \emph{{Naturalness priors and fits to the constrained
  minimal supersymmetric standard model}},
  \href{https://doi.org/10.1016/j.physletb.2006.02.052}{ \emph{Phys. Lett. B}
  {\bfseries 635} 123} [\href{https://arxiv.org/abs/hep-ph/0601089}{{\ttfamily
  hep-ph/0601089}}].

\bibitem[Allanach et~al.(2007)Allanach, Cranmer, Lester, and
  Weber]{allanach2007}
 Allanach~B.~C., Cranmer~K., Lester~C.~G. and Weber~A.~M., (2007).
  \emph{{Natural priors, CMSSM fits and LHC weather forecasts}},
  \href{https://doi.org/10.1088/1126-6708/2007/08/023}{ \emph{JHEP} {\bfseries
  08} 023} [\href{https://arxiv.org/abs/0705.0487}{{\ttfamily 0705.0487}}].

\bibitem[Anderson and Castano(1995)]{anderson1995}
 Anderson~G.~W. and Castano~D.~J., (1995). \emph{{Measures of fine tuning}},
  \href{https://doi.org/10.1016/0370-2693(95)00051-L}{ \emph{Phys. Lett. B}
  {\bfseries 347} 300} [\href{https://arxiv.org/abs/hep-ph/9409419}{{\ttfamily
  hep-ph/9409419}}].

\bibitem[Athron and Miller(2007)]{athron2007}
 Athron~P. and Miller~D.~J., (2007). \emph{{A New Measure of Fine Tuning}},
  \href{https://doi.org/10.1103/PhysRevD.76.075010}{ \emph{Phys. Rev. D}
  {\bfseries 76} 075010} [\href{https://arxiv.org/abs/0705.2241}{{\ttfamily
  0705.2241}}].

\bibitem[Athron et~al.(2017)Athron, Balazs, Farmer, Fowlie, Harries, and
  Kim]{athron2017}
 Athron~P., Balazs~C., Farmer~B., Fowlie~A., Harries~D. and Kim~D., (2017).
  \emph{{Bayesian analysis and naturalness of (Next-to-)Minimal Supersymmetric
  Models}}, \href{https://doi.org/10.1007/JHEP10(2017)160}{ \emph{JHEP}
  {\bfseries 10} 160} [\href{https://arxiv.org/abs/1709.07895}{{\ttfamily
  1709.07895}}].

\bibitem[Bain(2019)]{bain2019}
 Bain~J., (2019). \emph{{Why be Natural?}},
  \href{https://doi.org/10.1007/s10701-019-00249-z}{ \emph{Found. Phys.}
  {\bfseries 49} 898–914}.

\bibitem[Balasubramanian(1997)]{balasubramanian1997}
 Balasubramanian~V., (1997). \emph{{Statistical Inference, Occam's Razor, and
  Statistical Mechanics on the Space of Probability Distributions}},
  \href{https://doi.org/10.1162/neco.1997.9.2.349}{ \emph{Neural Comput.}
  {\bfseries 9} 349} [\href{https://arxiv.org/abs/cond-mat/9601030}{{\ttfamily
  cond-mat/9601030}}].

\bibitem[Balazs et~al.(2013)Balazs, Buckley, Carter, Farmer, and
  White]{balazs2013}
 Balazs~C., Buckley~A., Carter~D., Farmer~B. and White~M., (2013).
  \emph{{Should we still believe in constrained supersymmetry?}},
  \href{https://doi.org/10.1140/epjc/s10052-013-2563-y}{ \emph{Eur. Phys. J. C}
  {\bfseries 73} 2563} [\href{https://arxiv.org/abs/1205.1568}{{\ttfamily
  1205.1568}}].

\bibitem[Barbieri and Giudice(1988)]{Barbieri:1987fn}
 Barbieri~R. and Giudice~G.~F., (1988). \emph{{Upper Bounds on Supersymmetric
  Particle Masses}}, \href{https://doi.org/10.1016/0550-3213(88)90171-X}{
  \emph{Nucl. Phys. B} {\bfseries 306} 63}.

\bibitem[Borrelli and Castellani(2019)]{borrelli2019}
 Borrelli~A. and Castellani~E., (2019). \emph{{The Practice of Naturalness: A
  Historical-Philosophical Perspective}},
  \href{https://doi.org/10.1007/s10701-019-00287-7}{ \emph{Found. Phys.}
  {\bfseries 49} 860–878}.

\bibitem[Cabrera et~al.(2009)Cabrera, Casas, and Ruiz~de Austri]{cabrera2009}
 Cabrera~M.~E., Casas~J.~A. and Ruiz~de Austri~R., (2009). \emph{{Bayesian
  approach and Naturalness in MSSM analyses for the LHC}},
  \href{https://doi.org/10.1088/1126-6708/2009/03/075}{ \emph{JHEP} {\bfseries
  03} 075} [\href{https://arxiv.org/abs/0812.0536}{{\ttfamily 0812.0536}}].

\bibitem[Cabrera et~al.(2010)Cabrera, Casas, and Ruiz~de Austri]{cabrera2010}
 Cabrera~M.~E., Casas~J.~A. and Ruiz~de Austri~R., (2010). \emph{{MSSM Forecast
  for the LHC}}, \href{https://doi.org/10.1007/JHEP05(2010)043}{ \emph{JHEP}
  {\bfseries 05} 043} [\href{https://arxiv.org/abs/0911.4686}{{\ttfamily
  0911.4686}}].

\bibitem[Ciafaloni and Strumia(1997)]{Ciafaloni:1996zh}
 Ciafaloni~P. and Strumia~A., (1997). \emph{{Naturalness upper bounds on gauge
  mediated soft terms}}, \href{https://doi.org/10.1016/S0550-3213(97)00138-7}{
  \emph{Nucl. Phys. B} {\bfseries 494} 41}
  [\href{https://arxiv.org/abs/hep-ph/9611204}{{\ttfamily hep-ph/9611204}}].

\bibitem[Clarke and Cox(2017)]{clarke2017}
 Clarke~J.~D. and Cox~P., (2017). \emph{{Naturalness made easy: two-loop
  naturalness bounds on minimal SM extensions}},
  \href{https://doi.org/10.1007/JHEP02(2017)129}{ \emph{JHEP} {\bfseries 02}
  129} [\href{https://arxiv.org/abs/1607.07446}{{\ttfamily 1607.07446}}].

\bibitem[Cousins(2008)]{cousins2008}
 Cousins~R.~D., (2008). \emph{{Comment on `Bayesian Analysis of Pentaquark
  Signals from CLAS Data', with Response to the Reply by Ireland and
  Protopopsecu}}, \href{https://doi.org/10.1103/PhysRevLett.101.029101}{
  \emph{Phys. Rev. Lett.} {\bfseries 101} 029101}
  [\href{https://arxiv.org/abs/0807.1330}{{\ttfamily 0807.1330}}].

\bibitem[Craig(2023)]{craig2023}
 Craig~N., (2023). \emph{{Naturalness: past, present, and future}},
  \href{https://doi.org/10.1140/epjc/s10052-023-11928-7}{ \emph{Eur. Phys. J.
  C} {\bfseries 83} 825} [\href{https://arxiv.org/abs/2205.05708}{{\ttfamily
  2205.05708}}].

\bibitem[Crick(1989)]{crick1989}
 Crick~F., (1989).
\newblock \emph{{What mad pursuit: A personal view of scientific discovery}}.
\newblock Basic Books.

\bibitem[de~Carlos and Casas(1993)]{deCarlos:1993rbr}
 de~Carlos~B. and Casas~J.~A., (1993). \emph{{One loop analysis of the
  electroweak breaking in supersymmetric models and the fine tuning problem}},
  \href{https://doi.org/10.1016/0370-2693(93)90940-J}{ \emph{Phys. Lett. B}
  {\bfseries 309} 320} [\href{https://arxiv.org/abs/hep-ph/9303291}{{\ttfamily
  hep-ph/9303291}}].

\bibitem[de~Finetti(2017)]{definetti2017}
 de~Finetti~B., (2017).
\newblock \emph{{Theory of probability: A critical introductory treatment}}.
\newblock Wiley, 2nd edition,
  \href{https://doi.org/10.1002/9781119286387}{DOI}.

\bibitem[Dine(2015)]{dine2015}
 Dine~M., (2015). \emph{{Naturalness Under Stress}},
  \href{https://doi.org/10.1146/annurev-nucl-102014-022053}{ \emph{Ann. Rev.
  Nucl. Part. Sci.} {\bfseries 65} 43}
  [\href{https://arxiv.org/abs/1501.01035}{{\ttfamily 1501.01035}}].

\bibitem[Farmer(2015)]{farmer2015}
 Farmer~B., (2015).
\newblock \emph{{Epistemic probability and naturalness in global fits of
  supersymmetric models}}.
\newblock PhD thesis, Monash U.,
  \href{https://inspirehep.net/files/23cbbe0e5a78044fc7574bd420b78d80}{URL}.

\bibitem[Feng(2013)]{feng2013}
 Feng~J.~L., (2013). \emph{{Naturalness and the Status of Supersymmetry}},
  \href{https://doi.org/10.1146/annurev-nucl-102010-130447}{ \emph{Ann. Rev.
  Nucl. Part. Sci.} {\bfseries 63} 351}
  [\href{https://arxiv.org/abs/1302.6587}{{\ttfamily 1302.6587}}].

\bibitem[Fichet(2012)]{fichet2012}
 Fichet~S., (2012). \emph{{Quantified naturalness from Bayesian statistics}},
  \href{https://doi.org/10.1103/PhysRevD.86.125029}{ \emph{Phys. Rev. D}
  {\bfseries 86} 125029} [\href{https://arxiv.org/abs/1204.4940}{{\ttfamily
  1204.4940}}].

\bibitem[Fischer(2023)]{fischer2023}
 Fischer~E., (2023). \emph{{Naturalness and the Forward-Looking Justification
  of Scientific Principles}}, \href{https://doi.org/10.1017/psa.2023.5}{
  \emph{Philos. Sci.} {\bfseries 90} 1050–1059}.

\bibitem[Fischer(2024)]{fischer2024}
 Fischer~E., (2024). \emph{{Guiding principles in physics}},
  \href{https://doi.org/10.1007/s13194-024-00625-1}{ \emph{Eur. J. For Philos.
  Sci.} {\bfseries 14} }.

\bibitem[Fowlie(2013)]{fowlie2013}
 Fowlie~A., (2013).
\newblock \emph{{Bayesian Approach to Investigating Supersymmetric Models}}.
\newblock PhD thesis, Sheffield U.,
  \href{https://inspirehep.net/files/9827c4908285c6dae055d91d3eef1af8}{URL}.

\bibitem[Fowlie(2014{\natexlab{a}})]{fowlie2014a}
 Fowlie~A., (2014{\natexlab{a}}). \emph{{CMSSM, naturalness and the
  ``fine-tuning price'' of the Very Large Hadron Collider}},
  \href{https://doi.org/10.1103/PhysRevD.90.015010}{ \emph{Phys. Rev. D}
  {\bfseries 90} 015010} [\href{https://arxiv.org/abs/1403.3407}{{\ttfamily
  1403.3407}}].

\bibitem[Fowlie(2014{\natexlab{b}})]{fowlie2014b}
 Fowlie~A., (2014{\natexlab{b}}). \emph{{Is the CNMSSM more credible than the
  CMSSM?}}, \href{https://doi.org/10.1140/epjc/s10052-014-3105-y}{ \emph{Eur.
  Phys. J. C} {\bfseries 74} 3105}
  [\href{https://arxiv.org/abs/1407.7534}{{\ttfamily 1407.7534}}].

\bibitem[Fowlie(2024)]{fowlie2024}
 Fowlie~A., (2024). \emph{{The Bayes factor surface for searches for new
  physics}}, \href{https://doi.org/10.1140/epjc/s10052-024-12792-9}{ \emph{Eur.
  Phys. J. C} {\bfseries 84} 426}
  [\href{https://arxiv.org/abs/2401.11710}{{\ttfamily 2401.11710}}].

\bibitem[Fowlie and Herrera(2025)]{fowlie2025}
 Fowlie~A. and Herrera~G., (2025). \emph{{Precise interpretations of
  traditional fine-tuning measures}},
  \href{https://doi.org/10.1103/PhysRevD.111.015020}{ \emph{Phys. Rev. D}
  {\bfseries 111} 015020} [\href{https://arxiv.org/abs/2406.03533}{{\ttfamily
  2406.03533}}].

\bibitem[Fowlie et~al.(2016)Fowlie, Balazs, White, Marzola, and
  Raidal]{fowlie2016}
 Fowlie~A., Balazs~C., White~G., Marzola~L. and Raidal~M., (2016).
  \emph{{Naturalness of the relaxion mechanism}},
  \href{https://doi.org/10.1007/JHEP08(2016)100}{ \emph{JHEP} {\bfseries 08}
  100} [\href{https://arxiv.org/abs/1602.03889}{{\ttfamily 1602.03889}}].

\bibitem[Fundira and Purves(2018)]{fundira2018}
 Fundira~P. and Purves~A., (2018). \emph{{Bayesian naturalness, simplicity, and
  testability applied to the $B - L$ MSSM GUT}},
  \href{https://doi.org/10.1142/S0217751X1841004X}{ \emph{Int. J. Mod. Phys. A}
  {\bfseries 33} 1841004} [\href{https://arxiv.org/abs/1708.07835}{{\ttfamily
  1708.07835}}].

\bibitem[Gaillard and Lee(1974)]{Gaillard:1974hs}
 Gaillard~M.~K. and Lee~B.~W., (1974). \emph{{Rare Decay Modes of the K-Mesons
  in Gauge Theories}}, \href{https://doi.org/10.1103/PhysRevD.10.897}{
  \emph{Phys. Rev. D} {\bfseries 10} 897}.

\bibitem[Garrett(1991{\natexlab{a}})]{garrett1991a}
 Garrett~A., (1991{\natexlab{a}}). \emph{{Ockham's Razor}},
  \href{https://doi.org/10.1088/2058-7058/4/5/26}{ \emph{Phys. World}
  {\bfseries 4} 39}.

\bibitem[Garrett(1991{\natexlab{b}})]{garrett1991b}
 Garrett~A.
\newblock \emph{{Ockham's Razor}}, in Grandy~W.~T. and Schick~L.~H., editors,
  \emph{{Maximum Entropy and Bayesian Methods}}, pp.~357--364. Springer,
  Dordrecht, 1991{\natexlab{b}}.
\newblock ISBN 978-94-011-3460-6.

\bibitem[Gelman(2009)]{gelman2009}
 Gelman~A., (2009). \emph{{Bayes, Jeffreys, Prior Distributions and the
  Philosophy of Statistics}}, \href{https://doi.org/10.1214/09-sts284d}{
  \emph{Stat. Sci.} {\bfseries 24} }
  [\href{https://arxiv.org/abs/1001.2968}{{\ttfamily 1001.2968}}].

\bibitem[Gelman and Rubin(1995)]{gelman1995}
 Gelman~A. and Rubin~D.~B., (1995). \emph{{Avoiding Model Selection in Bayesian
  Social Research}}, \href{https://doi.org/10.2307/271064}{ \emph{Sociological
  Methodology} {\bfseries 25} 165}.

\bibitem[Giudice(2008)]{giudice2008}
 Giudice~G.~F.
\newblock \emph{{Naturally Speaking: The Naturalness Criterion and Physics at
  the LHC}}, in Kane~G. and Pierce~A., editors, \emph{{Perspectives on LHC
  Physics}}, pp.~155--178. World Scientific, January 2008,
  [\href{https://arxiv.org/abs/0801.2562}{{\ttfamily 0801.2562}}].
\newblock ISBN 9789812779762.

\bibitem[Giusti et~al.(1999)Giusti, Romanino, and Strumia]{giusti1999}
 Giusti~L., Romanino~A. and Strumia~A., (1999). \emph{{Natural ranges of
  supersymmetric signals}},
  \href{https://doi.org/10.1016/S0550-3213(99)00153-4}{ \emph{Nucl. Phys. B}
  {\bfseries 550} 3} [\href{https://arxiv.org/abs/hep-ph/9811386}{{\ttfamily
  hep-ph/9811386}}].

\bibitem[Good(1968)]{good1968}
 Good~I.~J., (1968). \emph{{Corroboration, Explanation, Evolving Probability,
  Simplicity and a Sharpened Razor}},
  \href{http://www.jstor.org/stable/686791}{ \emph{Br. J. Philos. Sci.}
  {\bfseries 19} 123}.

\bibitem[Good(1977)]{good1977}
 Good~I.~J., (1977). \emph{{Explicativity: a mathematical theory of explanation
  with statistical applications}},
  \href{https://doi.org/10.1098/rspa.1977.0069}{ \emph{Proc. R. Soc. London. A.
  Math. Phys. Sci.} {\bfseries 354} 303–330}.

\bibitem[Gregory(2005)]{gregory2005}
 Gregory~P., (2005).
\newblock \emph{Bayesian logical data analysis for the physical sciences}.
\newblock Cambridge University Press.

\bibitem[Grinbaum(2012)]{grinbaum2012}
 Grinbaum~A., (2012). \emph{{Which fine-tuning arguments are fine?}},
  \href{https://doi.org/10.1007/s10701-012-9629-9}{ \emph{Found. Phys.}
  {\bfseries 42} 615} [\href{https://arxiv.org/abs/0903.4055}{{\ttfamily
  0903.4055}}].

\bibitem[Gull(1988)]{gull1988}
 Gull~S.~F., (1988).
\newblock \emph{{Bayesian Inductive Inference and Maximum Entropy}}, in
  \emph{{Maximum Entropy and Bayesian Methods}}, p.~53–74.
\newblock Springer, \href{https://doi.org/10.1007/978-94-009-3049-0_4}{DOI},
  \href{https://bayes.wustl.edu/sfg/why.pdf}{URL}.

\bibitem[Hergt et~al.(2021)Hergt, Handley, Hobson, and Lasenby]{hergt2021}
 Hergt~L.~T., Handley~W.~J., Hobson~M.~P. and Lasenby~A.~N., (2021).
  \emph{{Bayesian evidence for the tensor-to-scalar ratio $r$ and neutrino
  masses $m_\nu$: Effects of uniform vs logarithmic priors}},
  \href{https://doi.org/10.1103/PhysRevD.103.123511}{ \emph{Phys. Rev. D}
  {\bfseries 103} 123511} [\href{https://arxiv.org/abs/2102.11511}{{\ttfamily
  2102.11511}}].

\bibitem[Hossenfelder(2019)]{hossenfelder2019}
 Hossenfelder~S., (2019). \emph{{Screams for explanation: finetuning and
  naturalness in the foundations of physics}},
  \href{https://doi.org/10.1007/s11229-019-02377-5}{ \emph{Synthese} {\bfseries
  198} 3727–3745} [\href{https://arxiv.org/abs/1801.02176}{{\ttfamily
  1801.02176}}].

\bibitem[Jaynes(1979)]{jaynes1979}
 Jaynes~E.~T., (1979). \emph{{\textup{Review of} Inference, Method, and
  Decision: Towards a Bayesian Philosophy of Science \textup{by
  R.D.~Rosenkrantz}}}, \href{https://doi.org/10.1080/01621459.1979.10481677}{
  \emph{J. Am. Stat. Assoc.} {\bfseries 74} 740}.

\bibitem[Jefferys and Berger(1991)]{jefferys1991}
 Jefferys~W.~H. and Berger~J.~O., (1991). \emph{{Sharpening Occam's Razor on a
  Bayesian strop}},
  \href{https://www.stat.purdue.edu/docs/research/tech-reports/1991/tr91-44c.pdf}{
  \emph{Bull. Am. Astron. Soc.} {\bfseries 23} 1259}.

\bibitem[Jefferys and Berger(1992{\natexlab{a}})]{jefferys1992a}
 Jefferys~W.~H. and Berger~J.~O., (1992{\natexlab{a}}). \emph{{The application
  of robust Bayesian analysis to hypothesis testing and Occam's razor}},
  \href{https://doi.org/10.1007/BF02589047}{ \emph{J. Ital. Stat. Soc.}
  {\bfseries 1} 17}.

\bibitem[Jefferys and Berger(1992{\natexlab{b}})]{jefferys1992b}
 Jefferys~W.~H. and Berger~J.~O., (1992{\natexlab{b}}). \emph{{Ockham's Razor
  and Bayesian Analysis}}, \href{http://www.jstor.org/stable/29774559}{
  \emph{Am. Sci.} {\bfseries 80} 64}.

\bibitem[Jeffreys(1961)]{jeffreys1961}
 Jeffreys~H., (1961).
\newblock \emph{{The Theory of Probability}}.
\newblock Oxford University Press, 3rd edition.

\bibitem[Kass and Raftery(1995)]{kass1995}
 Kass~R.~E. and Raftery~A.~E., (1995). \emph{{Bayes factors}},
  \href{https://doi.org/10.1080/01621459.1995.10476572}{ \emph{J. Am. Stat.
  Assoc.} {\bfseries 90} 773}.

\bibitem[Kim et~al.(2014)Kim, Athron, Bal{\'a}zs, Farmer, and
  Hutchison]{kim2014}
 Kim~D., Athron~P., Bal{\'a}zs~C., Farmer~B. and Hutchison~E., (2014).
  \emph{{Bayesian naturalness of the CMSSM and CNMSSM}},
  \href{https://doi.org/10.1103/PhysRevD.90.055008}{ \emph{Phys. Rev. D}
  {\bfseries 90} 055008} [\href{https://arxiv.org/abs/1312.4150}{{\ttfamily
  1312.4150}}].

\bibitem[Koren(2020)]{koren2020}
 Koren~S., (2020).
\newblock \emph{{New Approaches to the Hierarchy Problem and their Signatures
  from Microscopic to Cosmic Scales}}.
\newblock PhD thesis, UC, Santa Barbara (main),
  \href{https://arxiv.org/abs/2009.11870}{{\ttfamily 2009.11870}}.

\bibitem[Lindley and Phillips(1976)]{lindley1976}
 Lindley~D.~V. and Phillips~L.~D., (1976). \emph{{Inference for a Bernoulli
  Process (a Bayesian View)}},
  \href{https://doi.org/10.1080/00031305.1976.10479154}{ \emph{Am. Stat.}
  {\bfseries 30} 112–119}.

\bibitem[Lindley(2006)]{lindley2006}
 Lindley~D.~V., (2006).
\newblock \emph{{Understanding uncertainty}}.
\newblock Wiley, 11th edition, \href{https://doi.org/10.1002/0470055480}{DOI}.

\bibitem[Loredo(1990)]{loredo1990}
 Loredo~T.~J., (1990).
\newblock \emph{{From Laplace to Supernova SN 1987A: Bayesian Inference in
  Astrophysics}}, in \emph{{Maximum Entropy and Bayesian Methods}},
  p.~81–142.
\newblock Springer, \href{https://doi.org/10.1007/978-94-009-0683-9_6}{DOI},
  \href{https://hosting.astro.cornell.edu/~loredo/bayes/L90-LaplaceToSN1987A-scan.pdf}{URL}.

\bibitem[Lotfi et~al.(2022)Lotfi, Izmailov, Benton, Goldblum, and
  Wilson]{lotfi2022}
 Lotfi~S., Izmailov~P., Benton~G., Goldblum~M. and Wilson~A.~G., (2022).
\newblock \emph{{{B}ayesian Model Selection, the Marginal Likelihood, and
  Generalization}}, in Chaudhuri~K., Jegelka~S., Song~L., Szepesvari~C., Niu~G.
  and Sabato~S., editors, \emph{{Proceedings of the 39th International
  Conference on Machine Learning}}, volume 162 of \emph{Proc. Mach. Learn.
  Res.}, pp.~14223--14247. PMLR,
  \href{https://proceedings.mlr.press/v162/lotfi22a.html}{URL}.

\bibitem[MacKay(1991)]{mackay1991}
 MacKay~D.~J.~C., (1991).
\newblock \emph{{Bayesian Model Comparison and Backprop Nets}}, in Moody~J.,
  Hanson~S. and Lippmann~R., editors, \emph{{Advances in Neural Information
  Processing Systems}}, volume~4. Morgan-Kaufmann,
  \href{https://proceedings.neurips.cc/paper_files/paper/1991/file/c3c59e5f8b3e9753913f4d435b53c308-Paper.pdf}{URL}.

\bibitem[MacKay(1992{\natexlab{a}})]{mackay1992a}
 MacKay~D.~J.~C., (1992{\natexlab{a}}). \emph{{Bayesian Interpolation}},
  \href{https://doi.org/10.1162/neco.1992.4.3.415}{ \emph{Neural Comput.}
  {\bfseries 4} 415}.

\bibitem[MacKay(1992{\natexlab{b}})]{mackay1992b}
 MacKay~D.~J.~C., (1992{\natexlab{b}}).
\newblock \emph{{Bayesian methods for adaptive models}}.
\newblock PhD thesis, California Institute of Technology,
  \href{https://resolver.caltech.edu/CaltechETD:etd-01042007-131447}{URL},
  \href{https://doi.org/10.7907/H3A1-WM07}{DOI}.

\bibitem[MacKay(1992{\natexlab{c}})]{mackay1992c}
 MacKay~D.~J.~C., (1992{\natexlab{c}}). \emph{{A Practical Bayesian Framework
  for Backpropagation Networks}},
  \href{https://doi.org/10.1162/neco.1992.4.3.448}{ \emph{Neural Comput.}
  {\bfseries 4} 448}.

\bibitem[MacKay(1992{\natexlab{d}})]{mackay1992d}
 MacKay~D.~J.~C., (1992{\natexlab{d}}). \emph{{Information-Based Objective
  Functions for Active Data Selection}},
  \href{https://doi.org/10.1162/neco.1992.4.4.590}{ \emph{Neural Comput.}
  {\bfseries 4} 590}.

\bibitem[MacKay(1992{\natexlab{e}})]{mackay1992e}
 MacKay~D.~J.~C., (1992{\natexlab{e}}). \emph{{The Evidence Framework Applied
  to Classification Networks}},
  \href{https://doi.org/10.1162/neco.1992.4.5.720}{ \emph{Neural Comput.}
  {\bfseries 4} 720}.

\bibitem[MacKay(2003)]{mackay2003}
 MacKay~D.~J.~C., (2003).
\newblock \emph{{Information theory, inference and learning algorithms}}.
\newblock Cambridge University Press,
  \href{http://www.inference.org.uk/mackay/itila/book.html}{URL}.

\bibitem[Martin(1998)]{martin1998}
 Martin~S.~P., (1998). \emph{{A Supersymmetry primer}},
  \href{https://doi.org/10.1142/9789812839657_0001}{ \emph{Adv. Ser. Direct.
  High Energy Phys.} {\bfseries 18} 1}
  [\href{https://arxiv.org/abs/hep-ph/9709356}{{\ttfamily hep-ph/9709356}}].

\bibitem[McFadden(2023)]{mcfadden2023}
 McFadden~J., (2023). \emph{{Razor sharp: The role of Occam's razor in
  science}}, \href{https://doi.org/https://doi.org/10.1111/nyas.15086}{
  \emph{Ann. New York Acad. Sci.} {\bfseries 1530} 8}.

\bibitem[Murnane(2019)]{murnane2019}
 Murnane~D.~T., (2019).
\newblock \emph{{The landscape of composite Higgs models}}.
\newblock PhD thesis, Adelaide U.,
  \href{https://inspirehep.net/files/0b2115b02a6b9fdbf42d1355dc7772e2}{URL}.

\bibitem[Murray and Ghahramani(2005)]{murray2005}
 Murray~I. and Ghahramani~Z.
\newblock \emph{{A note on the evidence and {B}ayesian {O}ccam's razor}}, .
\newblock Technical Report GCNU-TR 2005-003, Gatsby Computational Neuroscience
  Unit, University College London, 2005,
  \href{https://mlg.eng.cam.ac.uk/zoubin/papers/05occam/occam.pdf}{URL}.

\bibitem[Neal(1996)]{neal1996}
 Neal~R.~M., (1996).
\newblock \emph{{Bayesian Learning for Neural Networks}}.
\newblock PhD thesis, Toronto U.,
  \href{https://www.cs.toronto.edu/pub/radford/thesis.pdf}{URL}.

\bibitem[O'Hagan(2004)]{o'hagan2004}
 O'Hagan~T., (2004). \emph{{Dicing with the Unknown}},
  \href{https://doi.org/10.1111/j.1740-9713.2004.00050.x}{ \emph{Significance}
  {\bfseries 1} 132–133}.

\bibitem[Peskin(2025)]{peskin2025}
 Peskin~M.~E., (2025). \emph{{What is the Hierarchy Problem?}},
  \href{https://doi.org/10.1016/j.nuclphysb.2025.116971}{ \emph{Nucl. Phys. B}
  {\bfseries 1018} 116971} [\href{https://arxiv.org/abs/2505.00694}{{\ttfamily
  2505.00694}}].

\bibitem[Rasmussen and Ghahramani(2000)]{rasmussen2000}
 Rasmussen~C. and Ghahramani~Z., (2000).
\newblock \emph{{Occam's Razor}}, in Leen~T., Dietterich~T. and Tresp~V.,
  editors, \emph{{Advances in Neural Information Processing Systems}},
  volume~13. MIT Press,
  \href{https://proceedings.neurips.cc/paper_files/paper/2000/file/0950ca92a4dcf426067cfd2246bb5ff3-Paper.pdf}{URL}.

\bibitem[Richter(2006)]{richter2006}
 Richter~B., (2006). \emph{{Theory in particle physics: Theological speculation
  versus practical knowledge}}, { \emph{Phys. Today} {\bfseries 59} 8}.

\bibitem[Robert(2016)]{robert2015}
 Robert~C.~P., (2016). \emph{{The expected demise of the Bayes factor}},
  \href{https://doi.org/10.1016/j.jmp.2015.08.002}{ \emph{J. Math. Psychol.}
  {\bfseries 72} 33–37} [\href{https://arxiv.org/abs/1506.08292}{{\ttfamily
  1506.08292}}].

\bibitem[Robert et~al.(2009)Robert, Chopin, and Rousseau]{robert2009}
 Robert~C.~P., Chopin~N. and Rousseau~J., (2009). \emph{{Harold Jeffreys’s
  Theory of Probability Revisited}}, \href{https://doi.org/10.1214/09-sts284}{
  \emph{Stat. Sci.} {\bfseries 24} }
  [\href{https://arxiv.org/abs/0804.3173}{{\ttfamily 0804.3173}}].

\bibitem[Rosenkrantz(1977)]{rosenkrantz1977}
 Rosenkrantz~R.~D., (1977).
\newblock \emph{{Simplicity}}, in \emph{{Inference, Method and Decision:
  Towards a Bayesian Philosophy of Science}}, pp.~93--117.
\newblock Springer, \href{https://doi.org/10.1007/978-94-010-1237-9_5}{DOI}.

\bibitem[Shifman(2012)]{Shifman:2012na}
 Shifman~M., (2012). \emph{{Frontiers Beyond the Standard Model: Reflections
  and Impressionistic Portrait of the conferernce}},
  \href{https://doi.org/10.1142/S0217732312300431}{ \emph{Mod. Phys. Lett. A}
  {\bfseries 27} 1230043} [\href{https://arxiv.org/abs/1211.0004}{{\ttfamily
  1211.0004}}].

\bibitem[Sivia and Skilling(2006)]{sivia2006}
 Sivia~D. and Skilling~J., (2006).
\newblock \emph{Data analysis: a Bayesian tutorial}.
\newblock Oxford University Press.

\bibitem[Smith and Spiegelhalter(1980)]{smith1980}
 Smith~A.~F.~M. and Spiegelhalter~D.~J., (1980). \emph{{Bayes Factors and
  Choice Criteria for Linear Models}},
  \href{http://www.jstor.org/stable/2984964}{ \emph{J. R. Stat. Soc. B}
  {\bfseries 42} 213}.

\bibitem[Sober(2015)]{sober2015}
 Sober~E., (2015).
\newblock \emph{{Ockham's razors: a user's manual}}.
\newblock Cambridge University Press.

\bibitem[Spiegelhalter(2024)]{spiegelhalter2024}
 Spiegelhalter~D., (2024). \emph{{Why probability probably doesn’t exist (but
  it is useful to act like it does)}},
  \href{https://doi.org/10.1038/d41586-024-04096-5}{ \emph{Nature} {\bfseries
  636} 560–563}.

\bibitem[Strumia(1999)]{strumia1999}
 Strumia~A., (1999).
\newblock \emph{{Naturalness of supersymmetric models}}, in \emph{{34th
  Rencontres de Moriond: Electroweak Interactions and Unified Theories}},
  pp.~441--446, [\href{https://arxiv.org/abs/hep-ph/9904247}{{\ttfamily
  hep-ph/9904247}}].

\bibitem[Susskind(1979)]{Susskind:1978ms}
 Susskind~L., (1979). \emph{{Dynamics of Spontaneous Symmetry Breaking in the
  Weinberg-Salam Theory}}, \href{https://doi.org/10.1103/PhysRevD.20.2619}{
  \emph{Phys. Rev. D} {\bfseries 20} 2619}.

\bibitem['t~Hooft(1980)]{tHooft:1979rat}
 't~Hooft~G., (1980). \emph{{Naturalness, chiral symmetry, and spontaneous
  chiral symmetry breaking}},
  \href{https://doi.org/10.1007/978-1-4684-7571-5_9}{ \emph{NATO Sci. Ser. B}
  {\bfseries 59} 135}.

\bibitem[Tendeiro et~al.(2024)Tendeiro, Kiers, Hoekstra, Wong, and
  Morey]{tendeiro2024}
 Tendeiro~J.~N., Kiers~H.~A.~L., Hoekstra~R., Wong~T.~K. and Morey~R.~D.,
  (2024). \emph{{Diagnosing the Misuse of the Bayes Factor in Applied
  Research}}, \href{https://doi.org/10.1177/25152459231213371}{ \emph{Adv.
  Methods Prac. Psychol. Sci.} {\bfseries 7} 25152459231213371}.

\bibitem[Wagenmakers and Matzke(2023)]{wagenmakers2023}
 Wagenmakers~E.-J. and Matzke~D., (2023).
\newblock \emph{{Bayesian inference from the ground up: The theory of common
  sense}}.
\newblock JASP,
  \href{https://www.bayesianspectacles.org/free-course-book/}{URL}.

\bibitem[Wallace(2019)]{wallace2019}
 Wallace~D., (2019). \emph{{Naturalness and Emergence}},
  \href{https://doi.org/10.1093/monist/onz022}{ \emph{Monist} {\bfseries 102}
  499}.

\bibitem[Weisskopf(1939)]{weisskopf1939}
 Weisskopf~V.~F., (1939). \emph{{On the Self-Energy and the Electromagnetic
  Field of the Electron}}, \href{https://doi.org/10.1103/PhysRev.56.72}{
  \emph{Phys. Rev.} {\bfseries 56} 72}.

\bibitem[Wells(2019{\natexlab{a}})]{wells2019a}
 Wells~J.~D., (2019{\natexlab{a}}). \emph{{Finetuned Cancellations and
  Improbable Theories}}, \href{https://doi.org/10.1007/s10701-019-00254-2}{
  \emph{Found. Phys.} {\bfseries 49} 428}
  [\href{https://arxiv.org/abs/1809.03374}{{\ttfamily 1809.03374}}].

\bibitem[Wells(2019{\natexlab{b}})]{wells2019b}
 Wells~J.~D., (2019{\natexlab{b}}). \emph{{Naturalness, Extra-Empirical Theory
  Assessments, and the Implications of Skepticism}},
  \href{https://doi.org/10.1007/s10701-018-0220-x}{ \emph{Found. Phys.}
  {\bfseries 49} 991} [\href{https://arxiv.org/abs/1806.07289}{{\ttfamily
  1806.07289}}].

\bibitem[Wells(2025)]{wells2025}
 Wells~J.~D., (2025). \emph{{The Intrinsic and Extrinsic Hierarchy Problems}},
  [\href{https://arxiv.org/abs/2506.05472}{{\ttfamily 2506.05472}}].

\bibitem[Williams(2015)]{Williams:2015gxa}
 Williams~P., (2015). \emph{{Naturalness, the autonomy of scales, and the 125
  GeV Higgs}}, \href{https://doi.org/10.1016/j.shpsb.2015.05.003}{ \emph{Stud.
  Hist. Phil. Sci. B} {\bfseries 51} 82}.

\bibitem[Williams(2018)]{williams2018}
 Williams~P., (2018). \emph{{Two Notions of Naturalness}},
  \href{https://doi.org/10.1007/s10701-018-0229-1}{ \emph{Found. Phys.}
  {\bfseries 49} 1022–1050}.

\bibitem[Wilson(1971)]{Wilson:1970ag}
 Wilson~K.~G., (1971). \emph{{The Renormalization Group and Strong
  Interactions}}, \href{https://doi.org/10.1103/PhysRevD.3.1818}{ \emph{Phys.
  Rev. D} {\bfseries 3} 1818}.

\bibitem[Wrinch and Jeffreys(1921)]{wrinch1921}
 Wrinch~D. and Jeffreys~H., (1921). \emph{{On certain fundamental principles of
  scientific inquiry}}, \href{https://doi.org/10.1080/14786442108633773}{
  \emph{London Edinburgh Dublin Philos. Mag. J. Sci.} {\bfseries 42}
  369–390}.

\end{thebibliography}

\end{document}